\documentclass[%
notitlepage,
 amsmath,amssymb,
]{revtex4-2}

\usepackage{graphicx}
\usepackage{dcolumn}
\usepackage{bm}
\usepackage[T1]{fontenc} 

\begin{document}

\title{Single-layer silicon metalens for broadband achromatic focusing and wide field of view}

\author{Jian Cao} \email{jian.cao@universite-paris-saclay.fr}
\author{Sarra Salhi}
\author{Jonathan Peltier}
\author{Jean-René Coudevylle}
\author{Samson Edmond}
\author{Cédric Villebasse}
\author{Etienne Herth}
\author{Laurent Vivien}
\author{Carlos Alonso-Ramos}
\author{Daniele Melati}
\affiliation{Centre de Nanosciences et de Nanotechnologies, Université Paris-Saclay, CNRS, 91120 Palaiseau, France}


\begin{abstract}
Achieving simultaneous broadband achromatic focusing and a wide field of view remains a significant challenge for metalenses. In this work, we begin with a quadratic phase profile, enabling full field-of-view designs, and apply dispersion engineering to minimize variations of the focal length across wavelengths, thereby substantially reducing both longitudinal and transverse chromatic aberrations. This is accomplished using only the propagation phase in waveguide-like rectangular meta-atoms, without relying on geometric phase contributions. The fabricated singlet metalens experimentally demonstrates a field of view of 86°, along with a tenfold reduction in focal length variations with wavelength compared to a conventional quadratic metalens, achieving a measured relative shift as low as 1.3\% across the 1.5 \textmu{}m - 1.6 \textmu{}m range (limited by our experimental setup). This improvement also leads to a twofold increase in focusing efficiency relative to the reference metalens. These experimental results validate the effectiveness of our design strategy in simultaneously enhancing the operational bandwidth and field of view of metalenses. The demonstrated performance can directly benefit beam steering applications in the near-infrared wavelength range and provides a path toward achromatic, wide field-of-view metalenses in the visible range for imaging systems.
\end{abstract}

\maketitle


\section{Introduction}

Metasurfaces are nano-structured surfaces realized by arranging meta-atoms at sub-wavelength distances \cite{lalanne_design_1999,yu2011light,yu2014flat}. Depending on the design of the meta-atoms and their arrangement, metasurfaces can be used to implement a variety of traditional optical functionalities, such as lensing \cite{chen2018broadband}, polarization manipulation \cite{teng2019conversion}, beam steering \cite{aieta2012out, su2015active}, or holographic image projection \cite{balthasar2017metasurface}. 
The possibility to replace bulk optical lenses with metasurfaces (also referred to as metalenses) is of interest for a range of applications, e.g. imaging, LiDAR, or free space communication devices, where the reduction of the system footprint is paramount in achieving scaling and widespread usage.
Demonstrating metalenses with sufficiently high performance is however crucial. Specifically, operation over a broadband wavelength range and a wide field of view have been standing challenges for metalenses, especially when the two need to be obtained simultaneously.

Chromatic aberration in metalenses depends on the optical dispersion of the meta-atoms and several approaches have been proposed to correct it and achieve an achromatic behavior \cite{yu2014flat, khorasaninejad2015achromatic, mohammad2018broadband, wang2018broadband, fathnan2018bandwidth, shrestha2018broadband, fan2019broadband,chen2020flat, balli2020hybrid, chung2020high, li2021meta, li2022inverse, liu2022broadband, sun2022broadband, fan2023advance, chu2023design, hu2023asymptotic, pan20233d}. Multi-layer metalenses, possibly using meta-atoms with different heights and combined with phase plates, can be used to largely reduce chromatic aberration while maintaining a high focusing efficiency \cite{balli2020hybrid, pan20233d}, at the cost of high fabrication complexity. As an example, by exploiting this method, Balli et al. \cite{balli2020hybrid} demonstrated a hybrid achromatic metalens working from 1.0 \textmu{}m to 1.8 \textmu{}m wavelength with an average focusing efficiencies of 60\%. 
Alternatively, chromatic aberration can be compensated by carefully engineering the phase profile of a single-layer metalens. By minimizing the group delay and the group delay dispersion of the phase profile, the imparted phase delay can in principle be made frequency-independent, hence achieving an achromatic behavior. 
This approach has been demonstrated by tailoring the geometric parameters of resonant or waveguide meta-atoms and also combining meta-atom shape tuning with geometric phase control \cite{pan2022dielectric}. Chen et al. \cite{chen2018broadband} demonstrated an achromatic metalens working in the visible range with a relative focal shift of 9.3\% over a 200 nm wavelength range around a central wavelength of 530 nm.
Lastly, instead of making the phase profile of a single layer metalens frequency independent, it is also possible to engineer its dispersion to ensure that the focal distance of the metalens remains constant despite the frequency change. Wavefronts at different wavelengths passing through a metalens will then experience different phase delays, but all corresponding to the same focal distance. By exploiting this method, Shrestha et al. \cite{shrestha2018broadband} demonstrated an achromatic metalens with a relative focal shift of 2\% over a 200 nm wavelength range around a working wavelength of 1.6 µm.

Besides chromatic aberration, metalenses focusing performance can also be severely affected by off-axis aberration. For example, a metalens with a hyperbolic phase profile has a high focusing quality for on-axis wavefronts. However, focusing capabilities rapidly degrade even for moderately tilted illumination \cite{liang2019high, lassalle2021imaging}. The problem of designing wide field-of-view metalenses has hence been extensively explored in the literature \cite{arbabi2017planar, groever2017meta, pu2017nanoapertures, martins2020metalenses,engelberg2020near, chen2022planar, fan2020ultrawide, lassalle2021imaging, li2021super,yang2023wide, xie2023large, dong2025full}. Placing an aperture stop in front of a metalens can help increase its field of view \cite{shalaginov2020single}, but it also largely reduces transmission efficiency. 
Arrays of metalenses \cite{chen2022planar}, where each metalens is optimized to focus light from different angles, have also been exploited for enlarged fields of view. This method, however, increases footprint, design, and fabrication complexity. Likewise, doublet metalenses or multilayer metalenses \cite{arbabi2016miniature, groever2017meta} can achieve a large field of view, requiring however complex fabrication processes. Finally, a singlet metalens can achieve full field of view thanks to judiciously designed phase profiles, e.g., modified hyperbolic phase profiles based on optimized polynomials, spherical phase profiles \cite{liang2019high}, or more commonly quadratic phase profiles \cite{pu2017nanoapertures, martins2020metalenses, lassalle2021imaging,liu2022broadband}. This approach usually trades off a wider field of view with a reduced focusing efficiency but ensures at the same time a relatively simple fabrication process and a low design complexity.

\begin{figure}[t]
    \centering
    \includegraphics[width=0.7\linewidth]{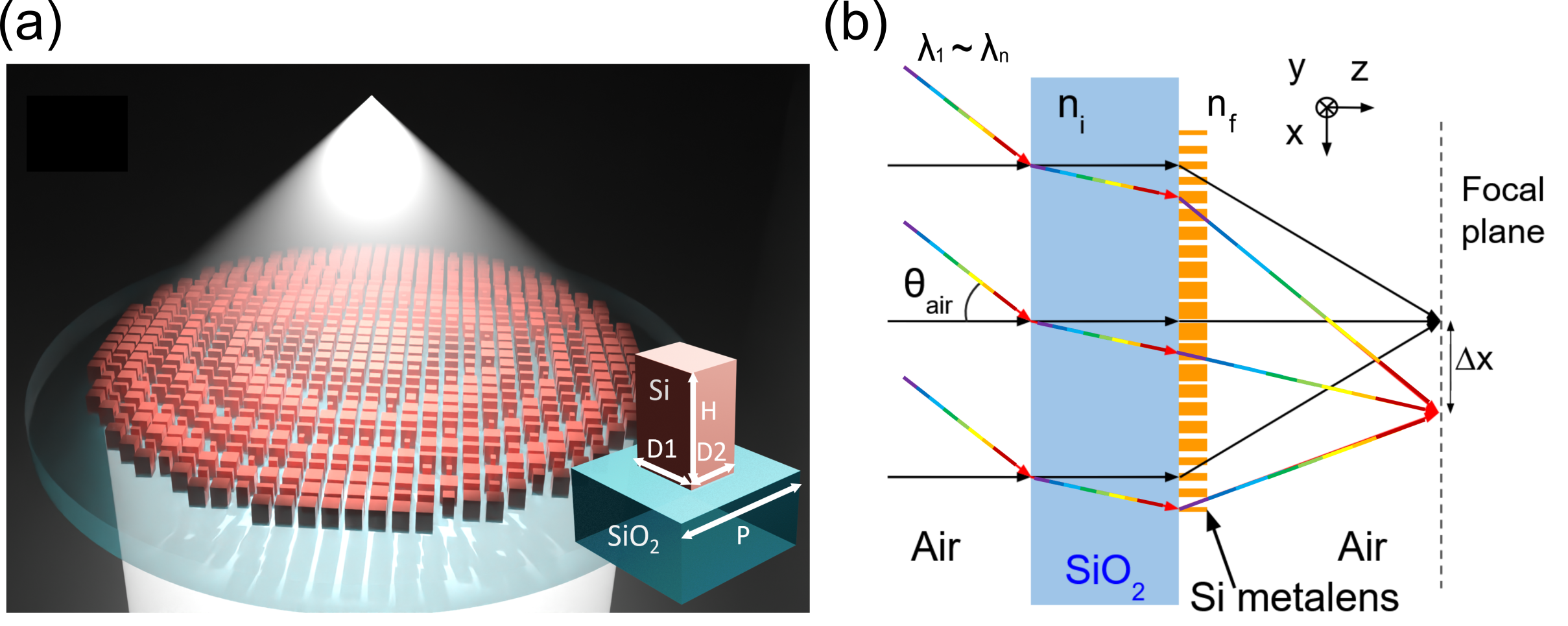}
    \caption{Achromatic and wide field of view metalens. (a) Pictorial representation of the metalens operation. In the inset, a sketch of the silicon meta-atoms is used to realize the metalens. The height is H = 700 nm. The sizes D1 and D2 vary from 150 nm to 500 nm with a period p = 650 nm. (b) Representation of the desired functionality. Longitudinal chromatic aberration is minimized and the focal plane distance from the metalens is maintained constant within a given wavelength range. At the same time, when the incident light is tilted of an angle $\theta_{air}$ in the x-z plane, the focal distance does not change, and the focal spot shifts of an amount $\Delta x$ along the x-axis (wide filed of view). Variations of $\Delta x$ with wavelength (transversal chromatic aberration) are minimized.}
    \label{fig:MetalensModel} 
\end{figure}
One possible solution to design a metalens that achieves at the same time achromatic focusing and a wide field of view consists of engineering the dispersion of a base phase profile chosen for its wide field of view. Previous designs reported by Xu et al. \cite{xu2023broadband} required the use of geometric phase to implement a quadratic phase profile using meta-atoms with different rotations and then controlled the phase profile dispersion exploiting the propagation phase within the waveguide-like meta-atoms. Simulation results showed a field of view up to 160° working in the 3.8 - 4.2 \textmu{}m wavelength range. Relying on a geometric phase contribution, the metalens required left-handed circularly polarized incident light. With a similar approach, Yu et al. \cite{hongli2024broadband} combined two metalenses, one based on the propagation phase and the second on the geometric phase, to design a doublet metalens with a field of view of 68° working on a 640 nm - 820 nm wavelength range (simulation results). Alternatively, without starting from a pre-defined phase profile, Yang et al. \cite{Yang:21} combined a direct search algorithm with a deep learning model to design free-form meta-atoms and exploit their diverse dispersion behaviors to directly optimize the design of metalens with a field of view of 160° working on a 1.0 \textmu{}m - 1.2 \textmu{}m wavelength range (simulation results).

In this work, we experimentally demonstrated a singlet achromatic and wide field-of-view metalens designed by engineering the dispersion of a quadratic phase profile used as a base. We demonstrated that we can achieve this goal without relying on a geometric phase contribution but solely using the propagation phase of silicon waveguide-like meta-atoms, hence removing the requirement for circularly polarized light. Experimental characterizations of the metalenses showed a field of view of 86° combined with a ten-fold reduction in the dependence of the focal distance on wavelength compared to a reference quadratic metalens without dispersion engineering. The measured relative focal shift was as low as 1.3\% in the 1.5 µm - 1.6 µm wavelength range, limited by our available setup, where the metalens also exhibited up to a twofold increase in focusing efficiency. As shown in Table 1 of the supplementary information document, this is, up to our knowledge, the best experimental performance for an achromatic and wide field-of-view singlet metalens reported so far in the literature.

\section{Design}\label{sec:design}

\begin{figure}[t]
    \centering
    \includegraphics[width=0.7\linewidth]{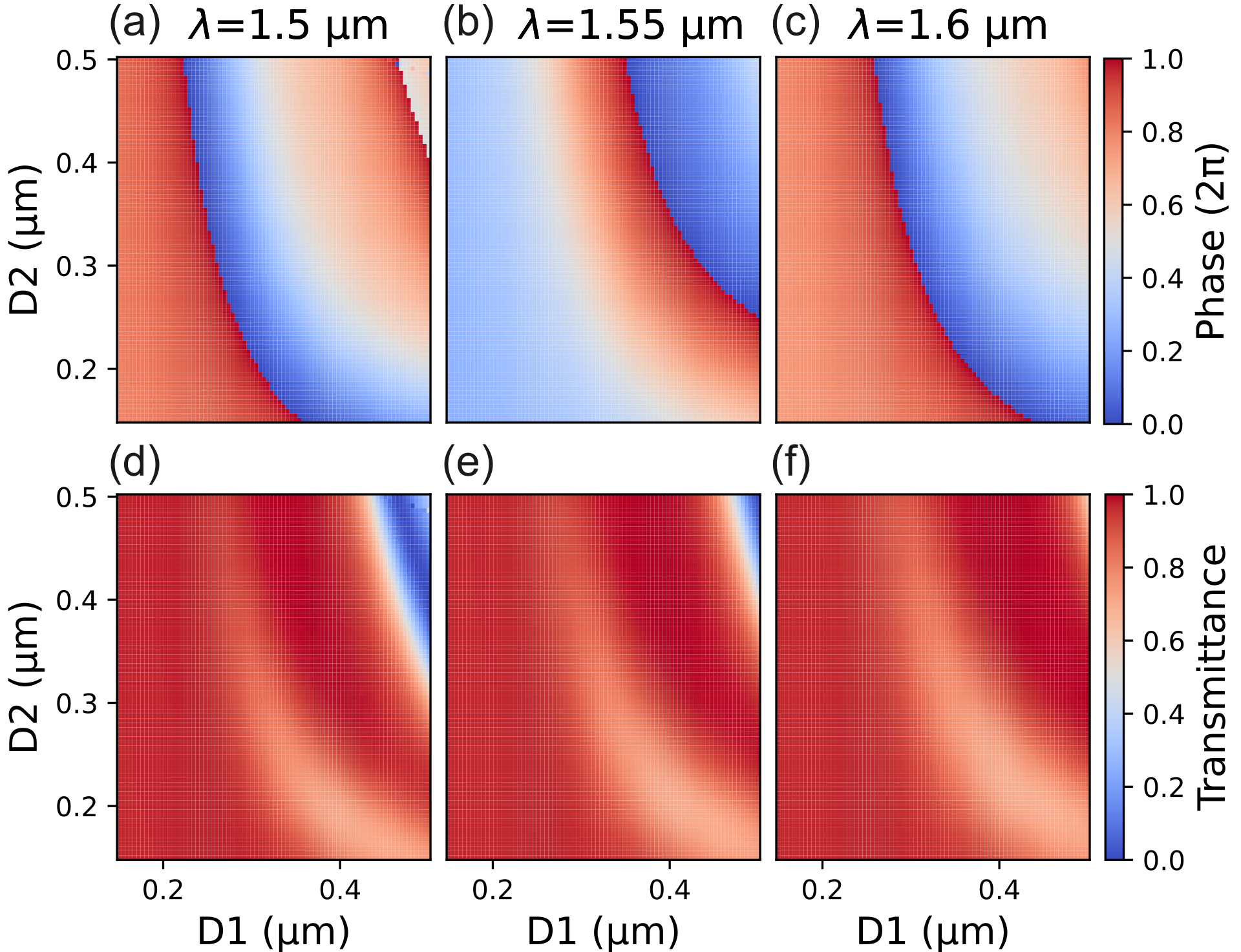}
    \caption{Phase delay and transmission efficiency of the meta-atoms. (a) - (c) Map of the phase delay imparted by meta-atoms with D1 and D2 varying from 150 nm to 500 nm and with a period of 650 nm, for three different wavelengths in the 1.5 µm - 1.6 µm range. In the simulations, light propagates from the $SiO_2$ toward the meta-atom. For each wavelength, the meta-atoms phase delay covers a 2$\pi$ range. (d) - (f) Corresponding transmission efficiency maps for the same series of meta-atoms. Most meta-atoms have an efficiency higher than 0.5 for the three wavelengths.}
    \label{fig:PhaseEfficiency}
\end{figure}

For the design of the metalens we considered 700-nm-thick silicon pillars with a rectangular shape placed on a SiO$_2$ substrate, as shown in Fig. \ref{fig:MetalensModel}(a). We exploited rigorous coupled-wave analysis (RCWA) \cite{hugonin2021reticolo} to simulate the behavior of the meta-atoms, assuming a locally periodic structure \cite{chung2020high,ueno2024ai}.
This approximation holds well with waveguide-like meta-atoms \cite{wang2018broadband}. We also assumed the incident light was TE-polarized. We built a library of available meta-atoms by sweeping the sizes D1 and D2 of the pillar from 150 nm to 500 nm while keeping the meta-atoms period fixed at 650 nm. In the sweeps, we limited the smallest feature (silicon pillars and pillar gaps) to be larger than 150 nm to improve fabrication reliability. We computed the phase delay and transmission efficiency imparted by each meta-atom shape to a plane wave propagating orthogonally to the silicon pillars for five different wavelengths in the 1.5 µm - 1.6 µm range. Simulation results are shown in Fig. \ref{fig:PhaseEfficiency} as a function of D1 and D2 for $\lambda$ = 1.5 µm, $\lambda$ = 1.55 µm, and $\lambda$ = 1.6 µm. Most of the meta-atoms efficiencies are higher than 50\% while their phase delay can cover the required 2$\pi$ range across the entire wavelength range. 

The design of the metalens was based on two objectives, as shown in Fig. \ref{fig:MetalensModel}(b). First, light coming at different angles needed to be focused on the same focal plane, with a fixed focal distance (wide field of view). At the same time, light at different wavelengths within a given range needed to be focused at the same focal spot, whose position was dictated by the angle of incidence (low longitudinal and transverse chromatic aberrations). We chose as a starting point for the design a quadratic phase profile which had already been demonstrated enabling focusing over a 180-degree field of view \cite{pu2017nanoapertures, martins2020metalenses, lassalle2021imaging, yang2023wide}:
\begin{equation} \label{eq:phaseprofile}
\begin{split}
    \phi(r, \lambda)\ &=\ -\ k\frac{r^2}{2f}\ =\ -\frac{\pi n_{f} r^2}{\lambda f}. 
    \end{split}
\end{equation}
In \eqref{eq:phaseprofile}, \textit{$\phi$}(r, $\lambda$) is the phase delay imparted by the metalens, which depends on the radial position r and wavelength $\lambda$, $k=n_fk_0$ is the wave-number in the propagation medium, $k_0$ is the wave-number in vacuum ($k_0=2\pi/\lambda$), $n_{f}$ is the refractive index in the focusing region, and $f$ is the designed focal length.

As schematically represented in Fig. \ref{fig:MetalensModel}(b), with a quadratic phase profile the focal spot shifts along the focal plane upon tilting of the incident beam. Assuming the incident beam is propagating along the z-axis and it is titled in the x-z plane with an angle $\theta_i$ from the normal to the metalens surface, the phase of the wavefront after the metalens is
\begin{align} 
\phi(r, \lambda)\ & =\ \phi_0\ -\ k\frac{r^2}{2f}\ -\ kxn_isin\theta_i \label{eq:Titltdphaseprofile} \\
& =\ \phi_0\ -\ \frac{\pi n_{f}}{\lambda f}(x^2 + y^2)\ -\ \frac{2\pi n_in_{f}}{\lambda}xsin\theta_i \nonumber\\
 \ &=\ \phi_0\ -\ \frac{\pi n_{f}}{\lambda f}[(x + n_i f sin\theta_i)^2 + y^2]\ +\ \frac{\pi n_{f} }{\lambda}n_i^2 f sin^2\theta_i \nonumber
\end{align}
where $n_i$ is the refractive index of the incident region. As shown in \eqref{eq:Titltdphaseprofile}, a tilt $\theta_i$ in the x-z plane introduces a spatial shift of the focal spot of $\Delta x = -n_ifsin\theta_i$ along the x-axis. The same is true for tilts in the orthogonal y-z plane, which causes a shift of the focal point along the y-axis. It should be noticed, however, that the phase delay in  \eqref{eq:phaseprofile} imparted by the metalens depends on the wavelength $\lambda$ of the incident wavefront, causing the focal spot to move in the longitudinal direction z, i.e., changing the focal distance (longitudinal chromatic aberration). This also causes at the same time a shift of the focal spot in the transversal direction for tilted illumination (transversal chromatic aberration) since $\Delta x$ depends on wavelength via the focal distance $f$, as can be seen in \eqref{eq:Titltdphaseprofile}.

\begin{figure}[t]
\centering
\includegraphics[width=0.7\linewidth]{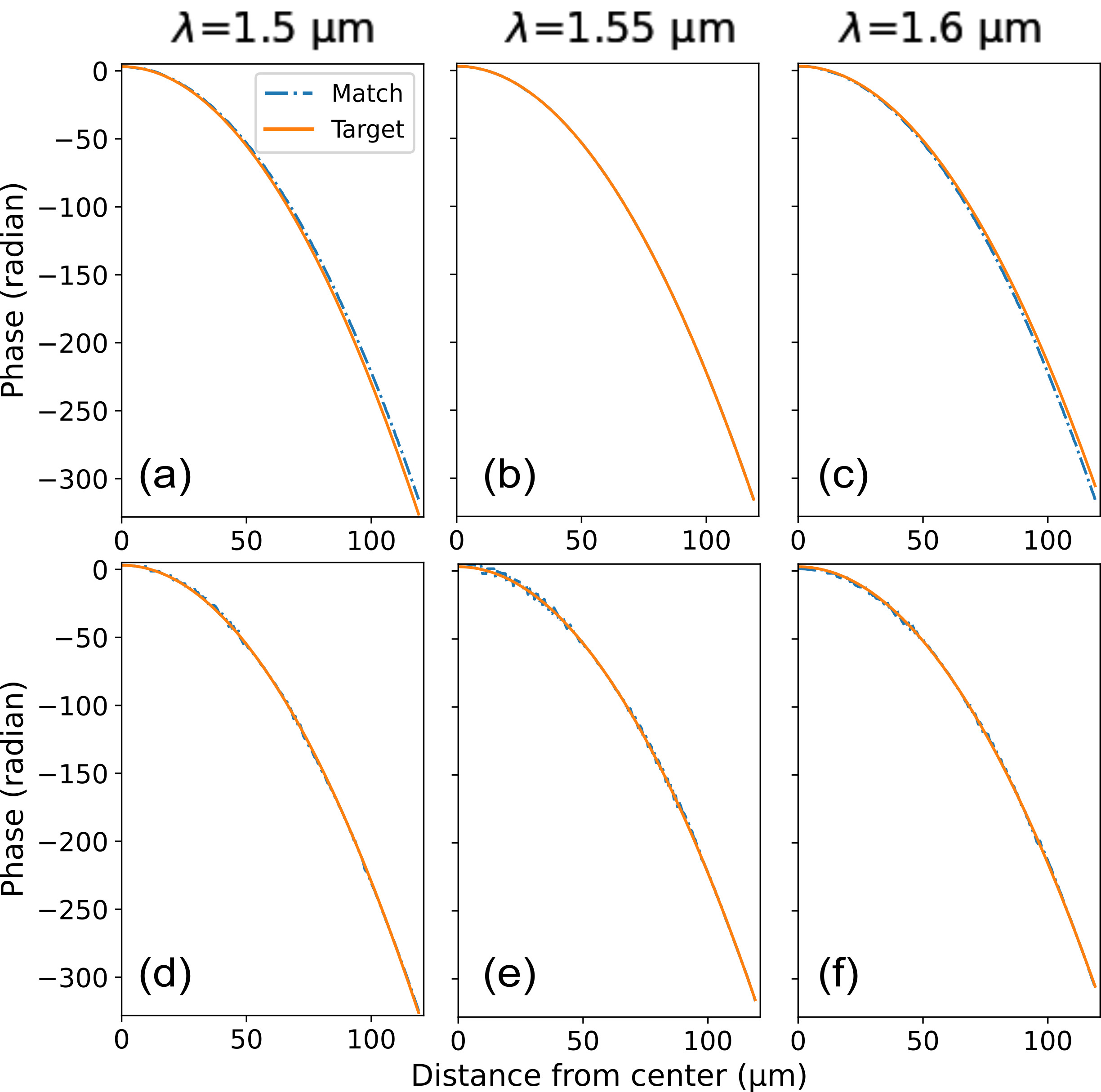}
\caption{Metalens phase profile matching. (a) - (c) Phase profile for a reference single-wavelength design metalens at three different wavelengths in the 1.5 µm - 1.6 µm range. At each position in the metasurface, we chose the meta-atom whose phase delay matched with the target value at $\lambda$ = 1.55 µm. The solid orange line is the target phase profile while the blue dotted line is the resulting matched phase profile. (d) - (f) Phase profile for the broadband design metalens for the same series of wavelengths. Instead of matching the required phase at only one wavelength, we chose the meta-atom which minimized the phase error at five different wavelengths in the 1.5 µm - 1.6 µm range}
\label{fig:PhaseProfile}
\end{figure}
\clearpage

In order to compensate for chromatic aberrations, we added as an additional design objective the fact that the metalens needed to match, at each wavelength, the correct target phase profile to ensure a fixed focal distance. We achieved this behavior by choosing within the library, for each position in the metalens, the meta-atom that minimized the difference between the target phase delay at that location and the actual phase delay of the meta-atom, averaged over five wavelengths in the 1.5 µm - 1.6 µm range:
\begin{equation}
    \begin{aligned}
    &(D_1(r), D_2(r))\ =\ (D_1,D_2)|_{min{\{\Delta\Phi\}}}\\  
    &with\ \Delta\Phi\ =\ E_i\{\Phi_{meta}(\lambda_i, r)\ -\ \Phi_{target}(\lambda_i,r)\}
    \label{eq:MiniPhase}
    \end{aligned}
\end{equation}
where $D_1$ and $D_2$ are the meta-atom sizes, $r$ is again the radial position from the center of metalens, $\lambda_i$ is the chosen wavelength, $\Delta\Phi$ is the difference between the meta-atom phase delay $\Phi_{meta}$ and the target phase $\Phi_{target}$, and $E_i$ denotes the average over wavelength. Throughout the manuscript, we refer to the metalens designed with this approach as the \textit{broadband design}.
As a reference, we also designed a regular wide field of view metalens operating at one single wavelength by matching the metalens phase profile only with the target quadratic phase at $\lambda$ =  1.55 µm.
We refer to this as the \textit{single-wavelength design}.

Figure \ref{fig:PhaseProfile} shows the obtained phase profile for a metalens with a diameter of 240 µm and a focal distance $f$ = 90 µm for both broadband and single-wavelength designs at three different wavelengths (dot-dashed blue lines). The figure also reports the quadratic phase profile that is required at each wavelength to obtain the desired focal distance of 90 µm (target profile, solid orange lines). For the single-wavelength design, the obtained phase profile matches well with the target requirement at $\lambda$ =  1.55 µm, but considerable deviations can be seen for the other wavelengths, see Fig. \ref{fig:PhaseProfile}(a-c). On the contrary, for the broadband design the obtained phase matches well with the target across the entire 1.5 µm - 1.6 µm spectrum, even if some additional noise can be seen around the center, Fig. \ref{fig:PhaseProfile}(d-f). This noise has an impact on the focusing performance, as discussed in Section \ref{sec:analysis}. Figure S1 of the supplementary information document reports directly the difference between the target and matched phase as a function of wavelength and position in the metalens for the two designs.

\section{Experimental Results}
\begin{figure}[t]
\centering
\includegraphics[width=0.7\linewidth]{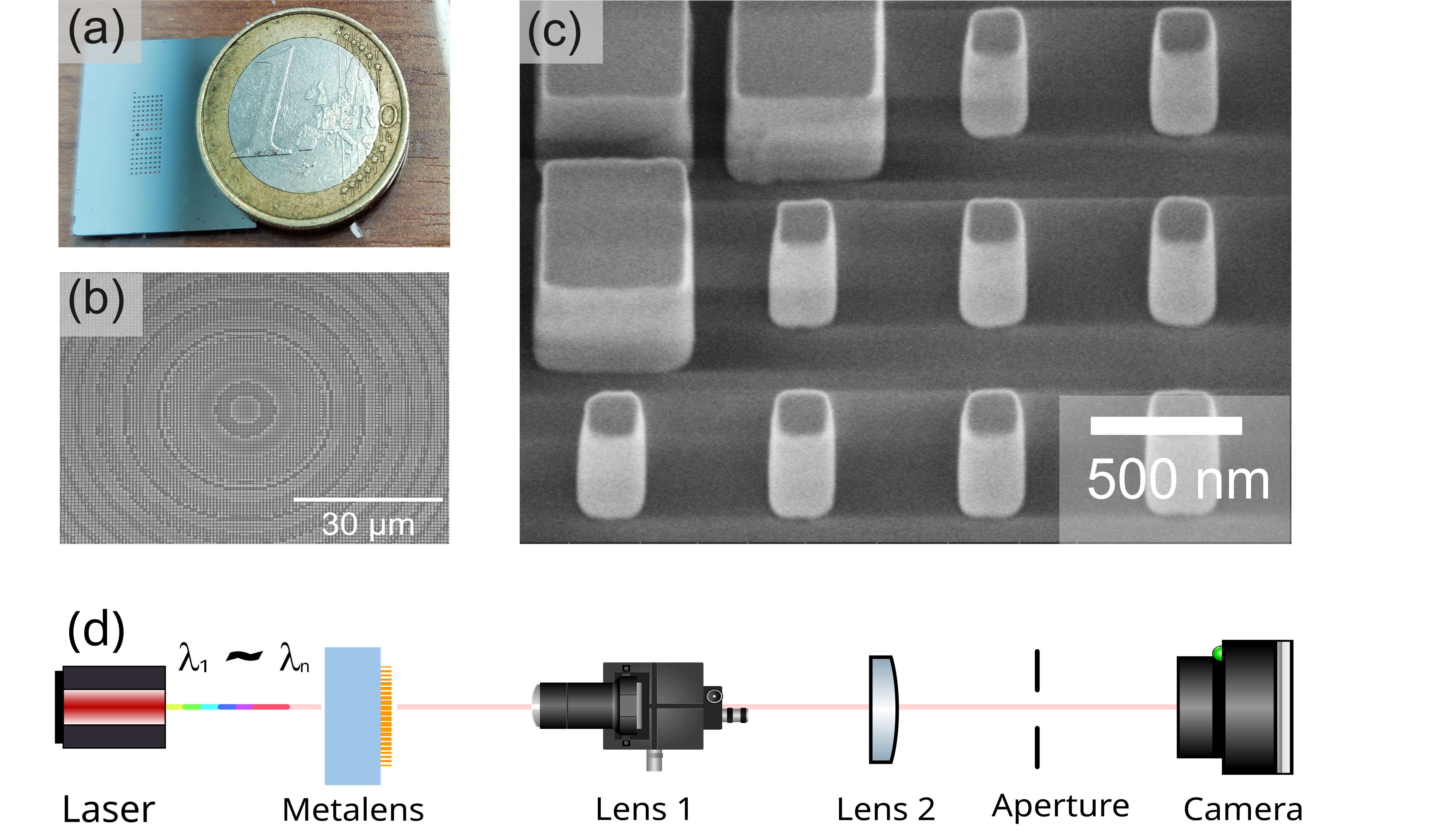}
\caption{Fabricated metalenses. (a) Optical image of a fabricated chip with several test metalenses. (b) An SEM image of the central part of one of the metalenses and (c) a zoom-in on the silicon meta-atoms. The smallest pillars are 150 nm x 150 nm in size. (d) Schematic of the experimental setup used for the characterization of the metalenses. The wavelength of the laser source could be tuned in the 1.5 µm - 1.6 µm range and its angular position with respect to the sample changed up to 90°.}
\label{fig:Experiments}
\end{figure}

Metalenses were fabricated using electron beam lithography and RIE - ICP etching (see Method). Optical and scanning electron microscope images of the fabricated devices are reported in Figs. \ref{fig:Experiments}(a-c). For both the broadband and the reference single wavelength designs, we fabricated different metalenses with numerical apertures NA of 0.74, 0.8, and 0.86, and a constant focal length of 90 µm. 

The setup shown in Fig.\ref{fig:Experiments}(d) was used to characterize the focal spot, the field of view, and the focal length of the fabricated metalenses (see Method). The laser beam passed through the metalens and the focal spot was imaged on the camera through an objective and corresponding tube lens. The wavelength of the laser was tuned between 1.5 µm and 1.6 µm and the illumination angle could be varied from 0° to 90°. The metalens position could be finely tuned through a piezoelectric stage. For the characterization of the focusing efficiency, a power meter was used instead of the camera.

\begin{figure*}[t]
\centering
\includegraphics[width=\linewidth]{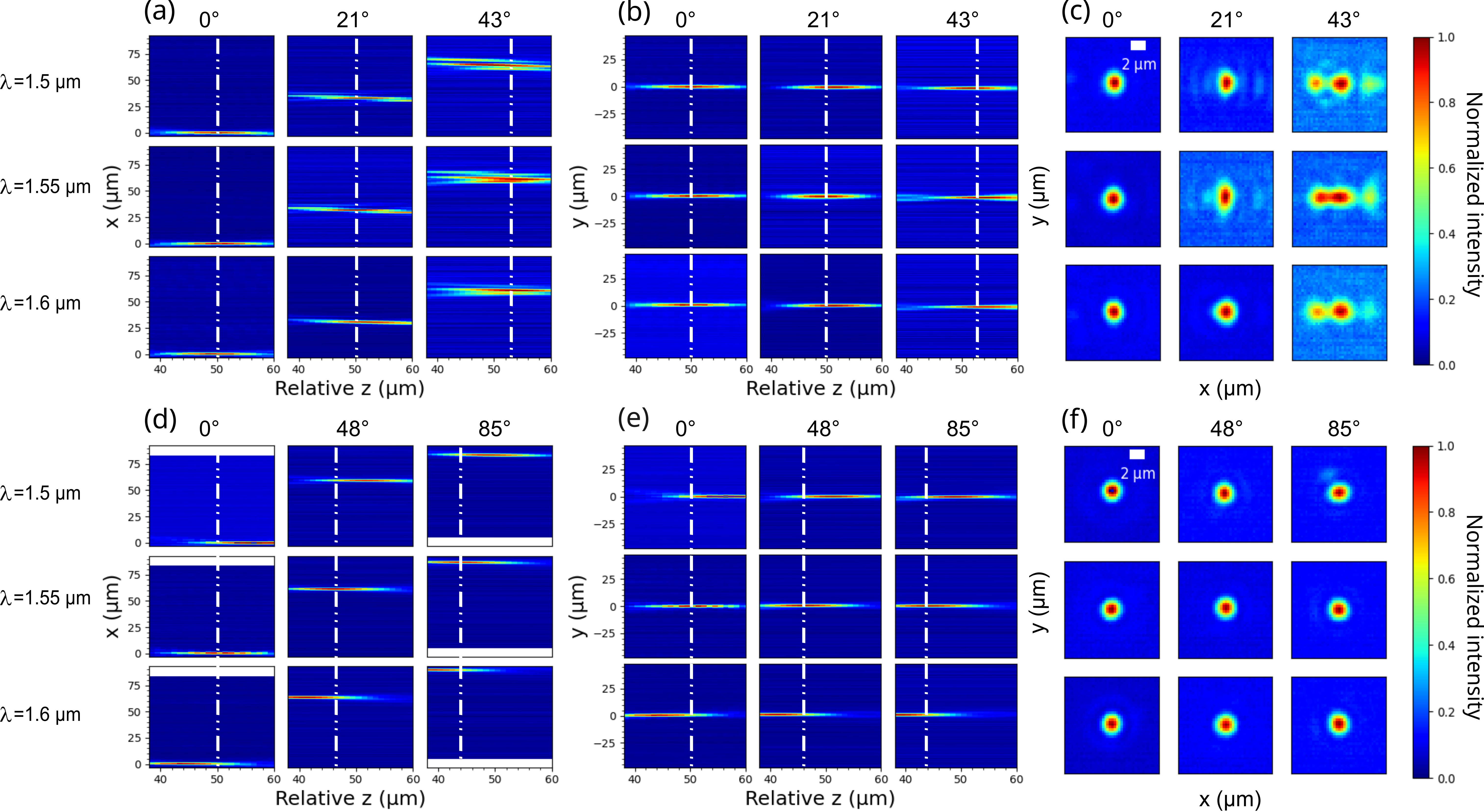}
\caption{Experimental characterization of the metalens focusing. (a-c) Results for the broadband metalens with NA = 0.8 and f = 90 um at $\lambda$ = 1.5 um, $\lambda$ = 1.55 um and $\lambda$ = 1.6 um. Panel (a) shows a cross-section of the focal spot in the x-z plane while (b) in the y-z plane, as defined in Fig. 1(b), with tilted incidence in in x-z plane from 0° to 43°. The white dashed lines mark the center of the focal spot at $\lambda$ = 1.55 µm. (c) Image of the focal spot in the x-y plane. (d) - (f) The same series of results for the single-wavelength metalens with NA = 0.8, and f = 90 µm. The incident angle varies from 0° to 85°.}
\label{fig:BroadbandPerform}
\end{figure*}

\smallskip

Figure \ref{fig:BroadbandPerform} shows the characterization of the focal spots of metalenses with NA = 0.8 and focal distance f= 90 µm at three different wavelengths in the 1.5 - 1.6 µm range and three different incident angles. The white dashed lines mark the focal spot position at $\lambda$ = 1.55 µm along the propagation axis z. Figures \ref{fig:BroadbandPerform}(a-c) refer to the broadband metalens, with incident angles of 0°, 21°, and 43°. Figures \ref{fig:BroadbandPerform}(d-f) report on the results for the single wavelength reference metalens with incident angles of 0°, 48°, and 85°. Figures \ref{fig:BroadbandPerform}(a,d) show a cross-section of the focusing pattern in the x-z plane, Figs. \ref{fig:BroadbandPerform}(b,e) in the y-z plane, and Figs. \ref{fig:BroadbandPerform}(c,f) in the x-y plane. 
As expected and discussed in Sec. \ref{sec:design}, both broadband and single wavelength metalenses show a transversal shift along the x-axis when the incident beam is tilted in the x-z plane. The broadband metalens has a field of view limited to $\pm$43° after which the focusing spot becomes severely degraded. The limitation on the field of view for the broadband design is discussed in detail in Sec. \ref{sec:analysis}. On the contrary, the single-wavelength metalens exhibit as expected a nearly full field of view (close to $\pm$90°), with almost no off-axis aberrations. However, in the considered wavelength range, the single-wavelength metalenses show a strong chromatic aberration, and the focal distance changes noticeably as the wavelength varies for all incident angles. For the broadband metalens, instead, the shift of the focal spot is largely reduced over the same wavelength range and for all the considered tilting angles.

\begin{figure}[t]
\centering
\includegraphics[width=0.7\linewidth]{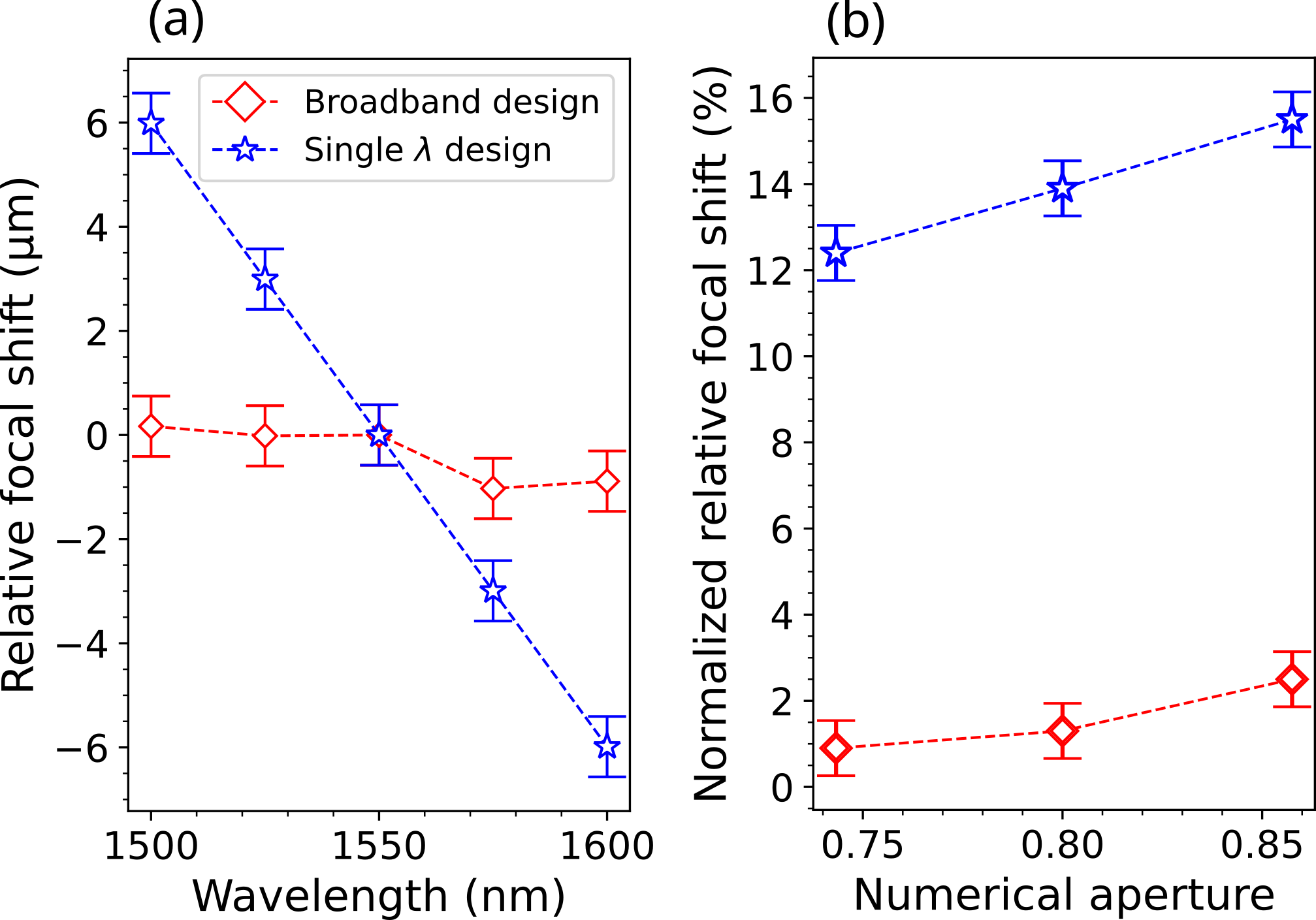}
\caption{Performance of the broadband and single wavelength metalens under normal illumination. (a) Relative shift of the focal distance at $\lambda$ = 1.55 µm for the broadband design (red diamond markers) and the single-wavelength metalens (blue star markers). (b) The normalized relative focal shift across the 100 nm bandwidth for the broadband design (red line) and the single-wavelength metalens (blue line)}
\label{fig:BroadbandSum}
\end{figure}
A more detailed analysis of the behavior of the two metalenses for normal illumination (0° tilting) is shown in Fig. \ref{fig:BroadbandSum}. 
In particular, Fig. \ref{fig:BroadbandSum}(a) reports on the variation of the focal distance as a function of the wavelength at $\lambda$ = 1.55 µm (relative focal shift) for broadband and single wavelength metalenses with NA = 0.8 and $f$ = 90 µm. The error bar is calculated as $\Delta u=\Delta\sigma / \sqrt{3}$, where $\Delta u$ is the measurement error and $\Delta\sigma$ is the measurement uncertainty of 1 µm coming from the experimental setup.
It can be seen that for the broadband design, the longitudinal chromatic aberration was largely compensated across the considered 100 nm wavelength bandwidth, with the relative focal shift changing less than 1.2 µm, i.e., 1.3 \% compared to the designed focal length. On the contrary, for the single-wavelength metalens, the relative focal shift exhibits a linear dependence on wavelength with a variation of 12.0 µm across the 100 nm bandwidth, ten times larger than the broadband metalenses.
Figure \ref{fig:BroadbandSum}(b) shows the dependence on the numerical aperture of the normalized relative focal shift, i.e.,  the ratio between the relative focal shift across the 100 nm bandwidth and the designed focal length. For the broadband design, the normalized relative focal shift remains smaller than 2 \% for numerical apertures between 0.75 and 0.85. For the regular single wavelength design, the normalized relative focal shifts are comprised between 12 \% at NA = 0.75 and 15 \% at NA = 0.85. For both broadband and single-wavelength designs, the normalized relative focal shift tends to worsen as the numerical aperture increases. This could be explained by the fact that the achromatic bandwidth decreases for large numerical apertures \cite{fathnan2018bandwidth,presutti2020focusing}.

\begin{figure}[t]
\centering
\includegraphics[width=0.7\linewidth]{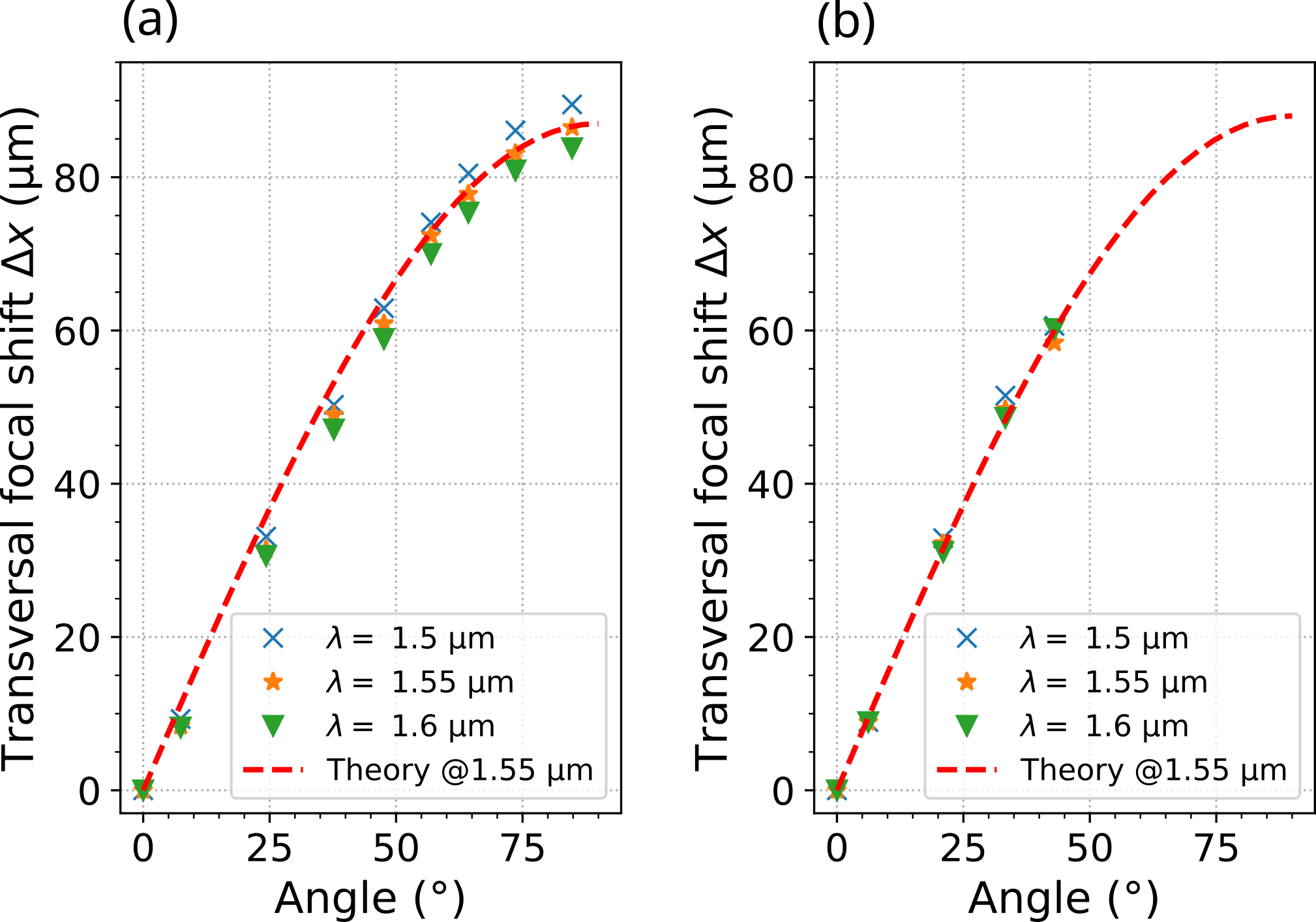}
\caption{Focal shift under tilted illumination. Figures report on the transversal focal shifts along the x-axis when the incidence angle is titled in the x-z plane for three different wavelengths in the 1.5 µm - 1.6 µm range, for (a) the single wavelength metalens and (b) the broadband design. Red dashed lines show the theoretical shift obtained by fitting results at $\lambda$ = 1.55 µm.}
\label{fig:WFOV}
\end{figure}

The characterization of the single wavelength and broadband metalenses under tilted illumination is reported in Fig. \ref{fig:WFOV} which shows the shift of the focal spot along the transversal direction x as a function of the incident angle for $\lambda$ = 1.5 µm, $\lambda$ = 1.55 µm and $\lambda$ = 1.6 µm. The transversal focal shift is calculated with respect to the position of the focal spot for an incident angle of 0°. As expected, as the incident angle increases, the single wavelength design shows larger variations of the transversal focal shifts across the 100 nm bandwidth compared to the broadband design (transversal chromatic aberration). For the single wavelength metalens, the difference in the transversal focal shift between $\lambda$ = 1.5 µm and $\lambda$ = 1.6 µm is 2.5 µm when the incident angle is 24° and it increases to 4.0 µm when the incident angle is 48°. For the broadband design, the difference is 1.8 µm when the incident angle is 21° and it only slightly increases to 2.2 µm when the incident angle increases to 43°, demonstrating an effective compensation also of the transversal chromatic aberration in the broadband design. As a reference, red dotted lines report on the theoretical focal shift $\Delta x=-n_ifsin\theta_i$ (as derived from Eq. \eqref{eq:Titltdphaseprofile}) obtained by fitting the experimental results at $\lambda$ = 1.55 µm. The fitted focal distances at this wavelength for the single wavelength and broadband designs were 87 µm and 88 µm, in good agreement with the designed focal length of 90 µm.

\begin{figure}[t]
\centering
\includegraphics[width=0.7\linewidth]{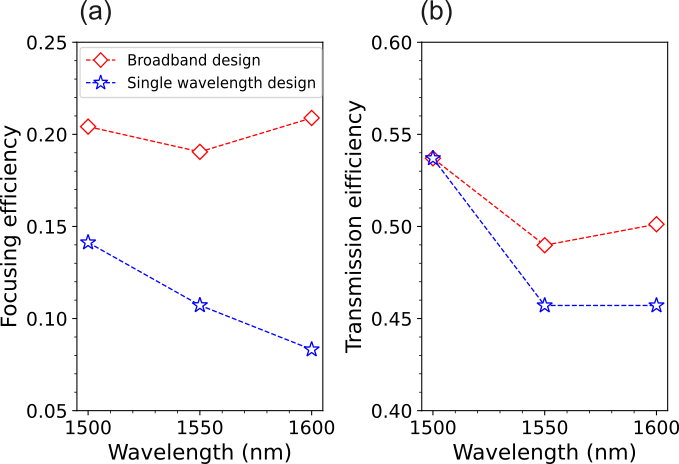}
\caption{Metalenses efficiency. (a) Focusing efficiency at normal incidence for the broadband (red line with diamond markers) and single-wavelength metalenses (blue line with star markers) with NA = 0.8 in the 1.5 µm - 1.6 µm wavelength range. (b) Transmission efficiency for the same metalenses.}
\label{fig:EnergyAspect}
\end{figure}

Lastly, we characterized the focusing and transmission efficiencies of both single wavelength and broadband metalenses, considering again NA = 0.8 and $f$ = 90 µm, over the considered 100 nm bandwidth and under normal illumination. Focusing efficiency was defined as the energy in the x-y plane at the focal distance integrated within a circle three times larger than the focal spot radius divided by the background signal where no metalens was present (but just the unpatterned substrate). The focal spot radius was measured to be around 2 µm, as shown in Fig. \ref{fig:BroadbandPerform}(c,f). The characterization results as a function of the wavelength are reported with red lines in Fig.\ref{fig:EnergyAspect}. Transmission efficiency was instead defined as the ratio between the energy integrated across the entire metalens area and the background signal, reported with blue lines in Fig.\ref{fig:EnergyAspect}. It can be seen that the broadband design (solid lines) shows a slightly higher focusing efficiency with a smaller dependence on wavelength compared to the single wavelength design. In the 1.5 - 1.6 µm wavelength range, the focusing efficiency varies from 19.1\% to 20.9\% for the broadband metalens, while it decreases from 14.1\% to 8.3\% for single wavelength metalens. This is expected because the single-wavelength design has a large longitudinal focal shift due to chromatic aberration, causing the energy integration to decrease as the wavefront goes out of focus as a consequence of a wavelength variation. Likewise, transmission efficiency varies from 49.0\% to 53.7\% for the broadband design and it decreases from 53.7\% to 45.7\% for the single wavelength design.

Table S1 of the supplementary information document compares these results with other experimental performances reported in the literature for achromatic and wide field-of-view metalenses. The metalens described in this work have an efficiency in line with previous results for metasurfaces based on quadratic phase profiles. On the contrary, it exhibits the best performance for singlet metalenses in terms of combined normalized relative focal shift and field of view.

\section{Analysis}
\label{sec:analysis}
\begin{figure}[t]
\centering
\includegraphics[width=0.7\linewidth]{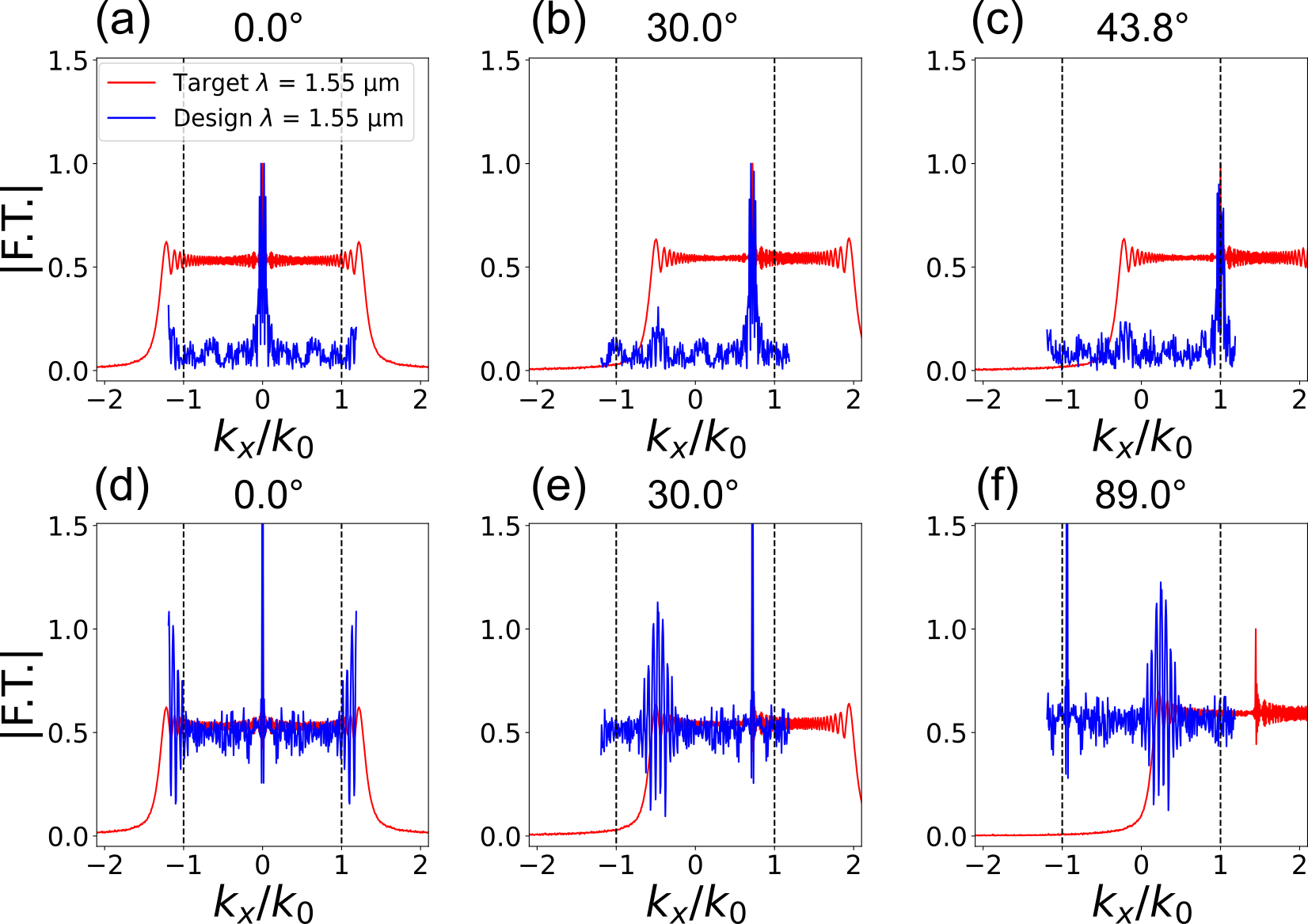}
\caption{Fourier transform spectra of the electric field after the metalenses along the normalized $k_x$ axis. Results for (a) - (c) the broadband design and (d) - (f) the single-wavelength design are shown with blue solid lines for three different illumination angles. Results for an ideal metalens exactly implementing the target phase profile at $\lambda$ = 1.55 \textmu{}m are shown with red solid lines. The black dotted lines mark the propagation region bounds in the normalized k-space.}
\label{fig:FTMix}
\end{figure}

For a more detailed analysis of the behavior of the designed metalenses, we focused on the broadband and single-wavelength designs with NA = 0.8 and f = 90 um at $\lambda$ = 1.55 µm, whose phase profiles are shown in Figs. \ref{fig:PhaseProfile}(e,b) (blue dashed line). We also considered an ideal metalens having exactly the target phase profile shown in Figs. \ref{fig:PhaseProfile}(e,b) with solid orange lines.
We assumed a plane wave illuminating the metalenses with different angles and we computed the distribution of the electric field right after the three metalenses described above. To this purpose, we considered both the phase delay and the transmission efficiency of each meta-atom as included in the prepared library and shown in Figs. \ref{fig:PhaseEfficiency}(b,e) at $\lambda$ = 1.55 µm. For the ideal metalens, we assumed that each meta-atom had unitary transmission efficiency. We then exploited Fourier analysis to study the outgoing fields. The resulting spectra in the spatial frequency domain for the three metalenses along the normalized $k_x/k_0$ axis (with $k_0$ the wave vector in vacuum) are reported in Fig. \ref{fig:FTMix}. Complete 2D spectra are shown in Fig. S2 of the supplementary information document. Figures \ref{fig:FTMix}(a-c) show with blue lines the normalized spatial spectrum for the broadband metalens for three different incident angles of 0°, 30°, and 43.8°. Figures \ref{fig:FTMix}(d-f) show with blue lines the same series of results but for the single-wavelength metalens for illuminations at 0°, 30°, and 89°. Normalized spatial frequencies for both broadband and single-wavelength metalenses are limited to $|k_x/k_0| < \pi/p/k_0 = 1.2$, determined by the metasurface period $p$ = 650 nm. Red lines in each panel report the results for the ideal metalens for the corresponding illumination angle.
Black vertical dashed lines mark the propagation region in k-space, defined as the range of spatial frequencies that can propagate in free space. Beyond this area, waves become evanescent with a purely imaginary z-axis wave vector $k_z$. The propagation condition can be written as:
\begin{equation}
    \sqrt{k_x^2 + k_y^2} < k,\ \text{with}\ k = n_f k_0
\end{equation}
where \( n_f \) is the refractive index in the focusing region (with \( n_f = 1 \) in our case). It should be noticed that the spatial spectrum of a metalens depends on its numerical aperture and in our case a value of NA = 0.8 or larger ensures non-zero components (at least in the ideal case) across the entire $k_x/k_0$ propagation range [-1,1] (for normal incidence), as it is shown also in Fig. S3 of the supplementary information document.

For the ideal metalens with an incident plane wave normal to the metasurface (Figs. \ref{fig:FTMix}(a,d), red lines) the normalized spatial spectrum has its maximum at $k_x=0$ and its Fourier components maintain as expected non-zero values across the entire [-1,1] range. This spatial spectrum closely resemble that of the single-wavelength metalens, Fig. \ref{fig:FTMix}(d), blue line. On the contrary, the spectrum of the broadband metalens, Fig. \ref{fig:FTMix}(a), blue line, is essentially concentered around $k_x=0$, with nearly zero amplitude for different $k_x$ values, despite the design numerical aperture was maintained at NA = 0.8. This difference can be attributed to the noisier phase profile that resulted from the design procedure described in Sec. \ref{sec:design} and that can be seen in Fig. \ref{fig:PhaseProfile}(e).

Tilting the angle of the incident plane wave causes the spectra to shift in the k-space as shown in Figs. \ref{fig:FTMix}(b,c,e,f), indicating a change of propagation direction that ultimately causes the transversal shift of the focal spot. When the angle of the incident wave is tilted by 30°, Figs. \ref{fig:FTMix}(b,e), spectra shift toward positive $k_x/k_0$ values of about $\Delta k_x(\theta) = n_i/n_f\ sin(\theta)$ = 0.72. Despite the shift, the single-wavelength metalens (as well as the ideal one) still maintain a largely non-zero spatial spectrum across the propagation region, ensuring the quality of the focal spot is maintained also with tilted illumination, see for example Fig. \ref{fig:BroadbandPerform}(f). The broadband metalens show instead a more asymmetrical spectrum, causing the emergence of stronger off-axis aberrations \cite{liang2019high} which tends to degrade the quality of the focal spot. This can be observed in Fig. \ref{fig:BroadbandPerform}(c) for an illumination tilting of 21°.

Finally, when the incident angle is larger than 43.8°, Fig. \ref{fig:FTMix}(c), the amplitude of the spatial spectrum for the broadband metalens within the propagation region becomes negligible. As a result, there is no more light propagating in the free space and the metalens reaches the limit of the field of view. This is what can be observed in the experimental results shown in Fig. \ref{fig:BroadbandPerform}(c) for an illumination tilting of 43°. The focal spot is deformed and could not be observed for larger illumination angles. As before, the broader spatial spectrum of the single-wavelength design ensures the focal spot is formed across the entire field of view up to a 90° tilting of the illuminating plane wave, Figs. \ref{fig:FTMix}(d) and \ref{fig:BroadbandPerform}(f). The difference between the spectrum of the ideal metalens and the single-wavelength one comes from the different sampling steps (650 nm for the single-wavelength metalens and 220 nm for the ideal phase profile), which cause different periodicity and aliasing in the Fourier domain. The impact on the Fourier spectrum of the sampling steps is further discussed in Fig. S4 of the supplementary information document.

\section{Conclusions}

In summary, we have demonstrated a single-layer, metalens with a broadband achromatic behavior and a wide field of view. By leveraging a library of waveguide-like silicon meta-atoms with diverse wavelength-dependent phase delays, we successfully engineered the dispersion of a quadratic phase profile without relying on geometric phase contributions and removing the requirement for circularly polarized incident light. These combined achievements represent a significant advancement over prior approaches, which often rely on doublet metalens designs or hybrid geometric–propagation phase methods. The experimental characterization across the 1.5 \textmu{}m – 1.6 \textmu{}m bandwidth (limited by our setup) showed focusing up to a field of view of $\pm$43° and a relative focal length shift as low as 1.3\%, an order-of-magnitude reduction compared to a conventional quadratic metalens used as a reference. As a consequence of the reduced chromatic aberration, focusing efficiency also improved up to a factor of two in the considered wavelength range compared to the reference metalens.

Furthermore, we exploited Fourier analysis to understand the impact of dispersion engineering on the metalens field of view. Our investigation suggests that the need to simultaneously match a target phase profile and its dispersion using the available library of meta-atoms introduced additional phase noise that shrunk the spatial spectra of the broadband metalens, causing a reduction of the field of view from the theoretical 180° limit. The limit predicted by the analysis ($\pm$43.8°) was indeed in very good agreement with the value measured experimentally. A possible way to address this limitation could be expanding the meta-atom design space using more diverse geometries, potentially guided by machine learning or optimization techniques. Such an approach could reduce phase noise, better match the target phase profile, and further improve both achromatic and field-of-view performance.

The combination of broadband achromaticity and wide angular performance that we have demonstrated, achieved through a single-layer design fully compatible with standard nanofabrication processes, offers a practical and scalable alternative to complex multi-layer architectures for integrating metalenses into photonic systems, particularly for beam steering applications in the near-infrared, offering also a promising pathway toward broadband and wide field of view metalenses for imaging systems operating in the visible range.

\section*{acknowledgement}

This work was partially funded by the European Union through the European Research Council (ERC) project BEAMS (Grant agreement No. 101041131). Views and opinions expressed are however those of the author(s) only and do not necessarily reflect those of the European Union or the European Research Council Executive Agency. Neither the European Union nor the granting authority can be held responsible for them. The device fabrication was performed within the C2N technological
platforms and partly supported by the RENATECH network and the General Council of Essonne.

\section*{Method}

\subsection{Metalenses fabrication} Metalenses were fabricated in the cleanroom of the Centre de Nanosciences et de Nanotechnologies using e-beam lithography with positive resist (ZEP) and proximity effect correction. After development (AR-600-546), structures were etched with inductively coupled plasma reactive ion etching (ICP RIE). The sample was finally cleaned using butanone, piranha solution, and oxygen plasma. 

\subsection{Determination of the focal distance} To determine the relative focal length of the metalenses, we first determined the focus x and y coordinated by searching for the maximum intensity. We then extracted the (normalized) intensity pattern along the z-axis at the focal spot position and we fitted it with a Gaussian function whose center we took as the z-coordinate of the focus. The relative focal length was then determined as the difference between the focal length at $\lambda$ = 1.55 µm and the focal length at the considered wavelength. It should be noticed that, for a given metalens, all the characterizations for different wavelengths and tilting angles have been performed without changing the position of the sample, hence ensuring the same origin for the axes.


\bibliography{broadbandMetalens}

\begin{thebibliography}{47}%
\makeatletter
\providecommand \@ifxundefined [1]{%
 \@ifx{#1\undefined}
}%
\providecommand \@ifnum [1]{%
 \ifnum #1\expandafter \@firstoftwo
 \else \expandafter \@secondoftwo
 \fi
}%
\providecommand \@ifx [1]{%
 \ifx #1\expandafter \@firstoftwo
 \else \expandafter \@secondoftwo
 \fi
}%
\providecommand \natexlab [1]{#1}%
\providecommand \enquote  [1]{``#1''}%
\providecommand \bibnamefont  [1]{#1}%
\providecommand \bibfnamefont [1]{#1}%
\providecommand \citenamefont [1]{#1}%
\providecommand \href@noop [0]{\@secondoftwo}%
\providecommand \href [0]{\begingroup \@sanitize@url \@href}%
\providecommand \@href[1]{\@@startlink{#1}\@@href}%
\providecommand \@@href[1]{\endgroup#1\@@endlink}%
\providecommand \@sanitize@url [0]{\catcode `\\12\catcode `\$12\catcode
  `\&12\catcode `\#12\catcode `\^12\catcode `\_12\catcode `\%12\relax}%
\providecommand \@@startlink[1]{}%
\providecommand \@@endlink[0]{}%
\providecommand \url  [0]{\begingroup\@sanitize@url \@url }%
\providecommand \@url [1]{\endgroup\@href {#1}{\urlprefix }}%
\providecommand \urlprefix  [0]{URL }%
\providecommand \Eprint [0]{\href }%
\providecommand \doibase [0]{https://doi.org/}%
\providecommand \selectlanguage [0]{\@gobble}%
\providecommand \bibinfo  [0]{\@secondoftwo}%
\providecommand \bibfield  [0]{\@secondoftwo}%
\providecommand \translation [1]{[#1]}%
\providecommand \BibitemOpen [0]{}%
\providecommand \bibitemStop [0]{}%
\providecommand \bibitemNoStop [0]{.\EOS\space}%
\providecommand \EOS [0]{\spacefactor3000\relax}%
\providecommand \BibitemShut  [1]{\csname bibitem#1\endcsname}%
\let\auto@bib@innerbib\@empty
\bibitem [{\citenamefont {Lalanne}\ \emph {et~al.}(1999)\citenamefont
  {Lalanne}, \citenamefont {Astilean}, \citenamefont {Chavel}, \citenamefont
  {Cambril},\ and\ \citenamefont {Launois}}]{lalanne_design_1999}%
  \BibitemOpen
  \bibfield  {author} {\bibinfo {author} {\bibfnamefont {P.}~\bibnamefont
  {Lalanne}}, \bibinfo {author} {\bibfnamefont {S.}~\bibnamefont {Astilean}},
  \bibinfo {author} {\bibfnamefont {P.}~\bibnamefont {Chavel}}, \bibinfo
  {author} {\bibfnamefont {E.}~\bibnamefont {Cambril}},\ and\ \bibinfo {author}
  {\bibfnamefont {H.}~\bibnamefont {Launois}},\ }\bibfield  {title} {\bibinfo
  {title} {Design and fabrication of blazed binary diffractive elements with
  sampling periods smaller than the structural cutoff},\ }\href
  {https://doi.org/10.1364/JOSAA.16.001143} {\bibfield  {journal} {\bibinfo
  {journal} {JOSA A}\ }\textbf {\bibinfo {volume} {16}},\ \bibinfo {pages}
  {1143} (\bibinfo {year} {1999})}\BibitemShut {NoStop}%
\bibitem [{\citenamefont {Yu}\ \emph {et~al.}(2011)\citenamefont {Yu},
  \citenamefont {Genevet}, \citenamefont {Kats}, \citenamefont {Aieta},
  \citenamefont {Tetienne}, \citenamefont {Capasso},\ and\ \citenamefont
  {Gaburro}}]{yu2011light}%
  \BibitemOpen
  \bibfield  {author} {\bibinfo {author} {\bibfnamefont {N.}~\bibnamefont
  {Yu}}, \bibinfo {author} {\bibfnamefont {P.}~\bibnamefont {Genevet}},
  \bibinfo {author} {\bibfnamefont {M.~A.}\ \bibnamefont {Kats}}, \bibinfo
  {author} {\bibfnamefont {F.}~\bibnamefont {Aieta}}, \bibinfo {author}
  {\bibfnamefont {J.-P.}\ \bibnamefont {Tetienne}}, \bibinfo {author}
  {\bibfnamefont {F.}~\bibnamefont {Capasso}},\ and\ \bibinfo {author}
  {\bibfnamefont {Z.}~\bibnamefont {Gaburro}},\ }\bibfield  {title} {\bibinfo
  {title} {Light propagation with phase discontinuities: generalized laws of
  reflection and refraction},\ }\href@noop {} {\bibfield  {journal} {\bibinfo
  {journal} {science}\ }\textbf {\bibinfo {volume} {334}},\ \bibinfo {pages}
  {333} (\bibinfo {year} {2011})}\BibitemShut {NoStop}%
\bibitem [{\citenamefont {Yu}\ and\ \citenamefont
  {Capasso}(2014)}]{yu2014flat}%
  \BibitemOpen
  \bibfield  {author} {\bibinfo {author} {\bibfnamefont {N.}~\bibnamefont
  {Yu}}\ and\ \bibinfo {author} {\bibfnamefont {F.}~\bibnamefont {Capasso}},\
  }\bibfield  {title} {\bibinfo {title} {Flat optics with designer
  metasurfaces},\ }\href@noop {} {\bibfield  {journal} {\bibinfo  {journal}
  {Nature materials}\ }\textbf {\bibinfo {volume} {13}},\ \bibinfo {pages}
  {139} (\bibinfo {year} {2014})}\BibitemShut {NoStop}%
\bibitem [{\citenamefont {Chen}\ \emph {et~al.}(2018)\citenamefont {Chen},
  \citenamefont {Zhu}, \citenamefont {Sanjeev}, \citenamefont {Khorasaninejad},
  \citenamefont {Shi}, \citenamefont {Lee},\ and\ \citenamefont
  {Capasso}}]{chen2018broadband}%
  \BibitemOpen
  \bibfield  {author} {\bibinfo {author} {\bibfnamefont {W.~T.}\ \bibnamefont
  {Chen}}, \bibinfo {author} {\bibfnamefont {A.~Y.}\ \bibnamefont {Zhu}},
  \bibinfo {author} {\bibfnamefont {V.}~\bibnamefont {Sanjeev}}, \bibinfo
  {author} {\bibfnamefont {M.}~\bibnamefont {Khorasaninejad}}, \bibinfo
  {author} {\bibfnamefont {Z.}~\bibnamefont {Shi}}, \bibinfo {author}
  {\bibfnamefont {E.}~\bibnamefont {Lee}},\ and\ \bibinfo {author}
  {\bibfnamefont {F.}~\bibnamefont {Capasso}},\ }\bibfield  {title} {\bibinfo
  {title} {A broadband achromatic metalens for focusing and imaging in the
  visible},\ }\href@noop {} {\bibfield  {journal} {\bibinfo  {journal} {Nature
  nanotechnology}\ }\textbf {\bibinfo {volume} {13}},\ \bibinfo {pages} {220}
  (\bibinfo {year} {2018})}\BibitemShut {NoStop}%
\bibitem [{\citenamefont {Teng}\ \emph {et~al.}(2019)\citenamefont {Teng},
  \citenamefont {Zhang}, \citenamefont {Wang}, \citenamefont {Liu},\ and\
  \citenamefont {Lv}}]{teng2019conversion}%
  \BibitemOpen
  \bibfield  {author} {\bibinfo {author} {\bibfnamefont {S.}~\bibnamefont
  {Teng}}, \bibinfo {author} {\bibfnamefont {Q.}~\bibnamefont {Zhang}},
  \bibinfo {author} {\bibfnamefont {H.}~\bibnamefont {Wang}}, \bibinfo {author}
  {\bibfnamefont {L.}~\bibnamefont {Liu}},\ and\ \bibinfo {author}
  {\bibfnamefont {H.}~\bibnamefont {Lv}},\ }\bibfield  {title} {\bibinfo
  {title} {Conversion between polarization states based on a metasurface},\
  }\href@noop {} {\bibfield  {journal} {\bibinfo  {journal} {Photonics
  Research}\ }\textbf {\bibinfo {volume} {7}},\ \bibinfo {pages} {246}
  (\bibinfo {year} {2019})}\BibitemShut {NoStop}%
\bibitem [{\citenamefont {Aieta}\ \emph {et~al.}(2012)\citenamefont {Aieta},
  \citenamefont {Genevet}, \citenamefont {Yu}, \citenamefont {Kats},
  \citenamefont {Gaburro},\ and\ \citenamefont {Capasso}}]{aieta2012out}%
  \BibitemOpen
  \bibfield  {author} {\bibinfo {author} {\bibfnamefont {F.}~\bibnamefont
  {Aieta}}, \bibinfo {author} {\bibfnamefont {P.}~\bibnamefont {Genevet}},
  \bibinfo {author} {\bibfnamefont {N.}~\bibnamefont {Yu}}, \bibinfo {author}
  {\bibfnamefont {M.~A.}\ \bibnamefont {Kats}}, \bibinfo {author}
  {\bibfnamefont {Z.}~\bibnamefont {Gaburro}},\ and\ \bibinfo {author}
  {\bibfnamefont {F.}~\bibnamefont {Capasso}},\ }\bibfield  {title} {\bibinfo
  {title} {Out-of-plane reflection and refraction of light by anisotropic
  optical antenna metasurfaces with phase discontinuities},\ }\href@noop {}
  {\bibfield  {journal} {\bibinfo  {journal} {Nano letters}\ }\textbf {\bibinfo
  {volume} {12}},\ \bibinfo {pages} {1702} (\bibinfo {year}
  {2012})}\BibitemShut {NoStop}%
\bibitem [{\citenamefont {Su}\ \emph {et~al.}(2015)\citenamefont {Su},
  \citenamefont {Ouyang}, \citenamefont {Xu}, \citenamefont {Cao},
  \citenamefont {Wei}, \citenamefont {Song}, \citenamefont {Gu}, \citenamefont
  {Tian}, \citenamefont {O’Hara}, \citenamefont {Han} \emph
  {et~al.}}]{su2015active}%
  \BibitemOpen
  \bibfield  {author} {\bibinfo {author} {\bibfnamefont {X.}~\bibnamefont
  {Su}}, \bibinfo {author} {\bibfnamefont {C.}~\bibnamefont {Ouyang}}, \bibinfo
  {author} {\bibfnamefont {N.}~\bibnamefont {Xu}}, \bibinfo {author}
  {\bibfnamefont {W.}~\bibnamefont {Cao}}, \bibinfo {author} {\bibfnamefont
  {X.}~\bibnamefont {Wei}}, \bibinfo {author} {\bibfnamefont {G.}~\bibnamefont
  {Song}}, \bibinfo {author} {\bibfnamefont {J.}~\bibnamefont {Gu}}, \bibinfo
  {author} {\bibfnamefont {Z.}~\bibnamefont {Tian}}, \bibinfo {author}
  {\bibfnamefont {J.~F.}\ \bibnamefont {O’Hara}}, \bibinfo {author}
  {\bibfnamefont {J.}~\bibnamefont {Han}}, \emph {et~al.},\ }\bibfield  {title}
  {\bibinfo {title} {Active metasurface terahertz deflector with phase
  discontinuities},\ }\href@noop {} {\bibfield  {journal} {\bibinfo  {journal}
  {Optics express}\ }\textbf {\bibinfo {volume} {23}},\ \bibinfo {pages}
  {27152} (\bibinfo {year} {2015})}\BibitemShut {NoStop}%
\bibitem [{\citenamefont {Balthasar~Mueller}\ \emph {et~al.}(2017)\citenamefont
  {Balthasar~Mueller}, \citenamefont {Rubin}, \citenamefont {Devlin},
  \citenamefont {Groever},\ and\ \citenamefont
  {Capasso}}]{balthasar2017metasurface}%
  \BibitemOpen
  \bibfield  {author} {\bibinfo {author} {\bibfnamefont {J.}~\bibnamefont
  {Balthasar~Mueller}}, \bibinfo {author} {\bibfnamefont {N.~A.}\ \bibnamefont
  {Rubin}}, \bibinfo {author} {\bibfnamefont {R.~C.}\ \bibnamefont {Devlin}},
  \bibinfo {author} {\bibfnamefont {B.}~\bibnamefont {Groever}},\ and\ \bibinfo
  {author} {\bibfnamefont {F.}~\bibnamefont {Capasso}},\ }\bibfield  {title}
  {\bibinfo {title} {Metasurface polarization optics: independent phase control
  of arbitrary orthogonal states of polarization},\ }\href@noop {} {\bibfield
  {journal} {\bibinfo  {journal} {Physical review letters}\ }\textbf {\bibinfo
  {volume} {118}},\ \bibinfo {pages} {113901} (\bibinfo {year}
  {2017})}\BibitemShut {NoStop}%
\bibitem [{\citenamefont {Khorasaninejad}\ \emph {et~al.}(2015)\citenamefont
  {Khorasaninejad}, \citenamefont {Aieta}, \citenamefont {Kanhaiya},
  \citenamefont {Kats}, \citenamefont {Genevet}, \citenamefont {Rousso},\ and\
  \citenamefont {Capasso}}]{khorasaninejad2015achromatic}%
  \BibitemOpen
  \bibfield  {author} {\bibinfo {author} {\bibfnamefont {M.}~\bibnamefont
  {Khorasaninejad}}, \bibinfo {author} {\bibfnamefont {F.}~\bibnamefont
  {Aieta}}, \bibinfo {author} {\bibfnamefont {P.}~\bibnamefont {Kanhaiya}},
  \bibinfo {author} {\bibfnamefont {M.~A.}\ \bibnamefont {Kats}}, \bibinfo
  {author} {\bibfnamefont {P.}~\bibnamefont {Genevet}}, \bibinfo {author}
  {\bibfnamefont {D.}~\bibnamefont {Rousso}},\ and\ \bibinfo {author}
  {\bibfnamefont {F.}~\bibnamefont {Capasso}},\ }\bibfield  {title} {\bibinfo
  {title} {Achromatic metasurface lens at telecommunication wavelengths},\
  }\href@noop {} {\bibfield  {journal} {\bibinfo  {journal} {Nano letters}\
  }\textbf {\bibinfo {volume} {15}},\ \bibinfo {pages} {5358} (\bibinfo {year}
  {2015})}\BibitemShut {NoStop}%
\bibitem [{\citenamefont {Mohammad}\ \emph {et~al.}(2018)\citenamefont
  {Mohammad}, \citenamefont {Meem}, \citenamefont {Shen}, \citenamefont
  {Wang},\ and\ \citenamefont {Menon}}]{mohammad2018broadband}%
  \BibitemOpen
  \bibfield  {author} {\bibinfo {author} {\bibfnamefont {N.}~\bibnamefont
  {Mohammad}}, \bibinfo {author} {\bibfnamefont {M.}~\bibnamefont {Meem}},
  \bibinfo {author} {\bibfnamefont {B.}~\bibnamefont {Shen}}, \bibinfo {author}
  {\bibfnamefont {P.}~\bibnamefont {Wang}},\ and\ \bibinfo {author}
  {\bibfnamefont {R.}~\bibnamefont {Menon}},\ }\bibfield  {title} {\bibinfo
  {title} {Broadband imaging with one planar diffractive lens},\ }\href@noop {}
  {\bibfield  {journal} {\bibinfo  {journal} {Scientific reports}\ }\textbf
  {\bibinfo {volume} {8}},\ \bibinfo {pages} {2799} (\bibinfo {year}
  {2018})}\BibitemShut {NoStop}%
\bibitem [{\citenamefont {Wang}\ \emph {et~al.}(2018)\citenamefont {Wang},
  \citenamefont {Wu}, \citenamefont {Su}, \citenamefont {Lai}, \citenamefont
  {Chen}, \citenamefont {Kuo}, \citenamefont {Chen}, \citenamefont {Chen},
  \citenamefont {Huang}, \citenamefont {Wang} \emph
  {et~al.}}]{wang2018broadband}%
  \BibitemOpen
  \bibfield  {author} {\bibinfo {author} {\bibfnamefont {S.}~\bibnamefont
  {Wang}}, \bibinfo {author} {\bibfnamefont {P.~C.}\ \bibnamefont {Wu}},
  \bibinfo {author} {\bibfnamefont {V.-C.}\ \bibnamefont {Su}}, \bibinfo
  {author} {\bibfnamefont {Y.-C.}\ \bibnamefont {Lai}}, \bibinfo {author}
  {\bibfnamefont {M.-K.}\ \bibnamefont {Chen}}, \bibinfo {author}
  {\bibfnamefont {H.~Y.}\ \bibnamefont {Kuo}}, \bibinfo {author} {\bibfnamefont
  {B.~H.}\ \bibnamefont {Chen}}, \bibinfo {author} {\bibfnamefont {Y.~H.}\
  \bibnamefont {Chen}}, \bibinfo {author} {\bibfnamefont {T.-T.}\ \bibnamefont
  {Huang}}, \bibinfo {author} {\bibfnamefont {J.-H.}\ \bibnamefont {Wang}},
  \emph {et~al.},\ }\bibfield  {title} {\bibinfo {title} {A broadband
  achromatic metalens in the visible},\ }\href@noop {} {\bibfield  {journal}
  {\bibinfo  {journal} {Nature nanotechnology}\ }\textbf {\bibinfo {volume}
  {13}},\ \bibinfo {pages} {227} (\bibinfo {year} {2018})}\BibitemShut
  {NoStop}%
\bibitem [{\citenamefont {Fathnan}\ and\ \citenamefont
  {Powell}(2018)}]{fathnan2018bandwidth}%
  \BibitemOpen
  \bibfield  {author} {\bibinfo {author} {\bibfnamefont {A.~A.}\ \bibnamefont
  {Fathnan}}\ and\ \bibinfo {author} {\bibfnamefont {D.~A.}\ \bibnamefont
  {Powell}},\ }\bibfield  {title} {\bibinfo {title} {Bandwidth and size limits
  of achromatic printed-circuit metasurfaces},\ }\href@noop {} {\bibfield
  {journal} {\bibinfo  {journal} {Optics express}\ }\textbf {\bibinfo {volume}
  {26}},\ \bibinfo {pages} {29440} (\bibinfo {year} {2018})}\BibitemShut
  {NoStop}%
\bibitem [{\citenamefont {Shrestha}\ \emph {et~al.}(2018)\citenamefont
  {Shrestha}, \citenamefont {Overvig}, \citenamefont {Lu}, \citenamefont
  {Stein},\ and\ \citenamefont {Yu}}]{shrestha2018broadband}%
  \BibitemOpen
  \bibfield  {author} {\bibinfo {author} {\bibfnamefont {S.}~\bibnamefont
  {Shrestha}}, \bibinfo {author} {\bibfnamefont {A.~C.}\ \bibnamefont
  {Overvig}}, \bibinfo {author} {\bibfnamefont {M.}~\bibnamefont {Lu}},
  \bibinfo {author} {\bibfnamefont {A.}~\bibnamefont {Stein}},\ and\ \bibinfo
  {author} {\bibfnamefont {N.}~\bibnamefont {Yu}},\ }\bibfield  {title}
  {\bibinfo {title} {Broadband achromatic dielectric metalenses},\ }\href@noop
  {} {\bibfield  {journal} {\bibinfo  {journal} {Light: Science \&
  Applications}\ }\textbf {\bibinfo {volume} {7}},\ \bibinfo {pages} {85}
  (\bibinfo {year} {2018})}\BibitemShut {NoStop}%
\bibitem [{\citenamefont {Fan}\ \emph {et~al.}(2019)\citenamefont {Fan},
  \citenamefont {Qiu}, \citenamefont {Zhang}, \citenamefont {Pang},
  \citenamefont {Zhou}, \citenamefont {Liu}, \citenamefont {Ren}, \citenamefont
  {Wang},\ and\ \citenamefont {Dong}}]{fan2019broadband}%
  \BibitemOpen
  \bibfield  {author} {\bibinfo {author} {\bibfnamefont {Z.-B.}\ \bibnamefont
  {Fan}}, \bibinfo {author} {\bibfnamefont {H.-Y.}\ \bibnamefont {Qiu}},
  \bibinfo {author} {\bibfnamefont {H.-L.}\ \bibnamefont {Zhang}}, \bibinfo
  {author} {\bibfnamefont {X.-N.}\ \bibnamefont {Pang}}, \bibinfo {author}
  {\bibfnamefont {L.-D.}\ \bibnamefont {Zhou}}, \bibinfo {author}
  {\bibfnamefont {L.}~\bibnamefont {Liu}}, \bibinfo {author} {\bibfnamefont
  {H.}~\bibnamefont {Ren}}, \bibinfo {author} {\bibfnamefont {Q.-H.}\
  \bibnamefont {Wang}},\ and\ \bibinfo {author} {\bibfnamefont {J.-W.}\
  \bibnamefont {Dong}},\ }\bibfield  {title} {\bibinfo {title} {A broadband
  achromatic metalens array for integral imaging in the visible},\ }\href@noop
  {} {\bibfield  {journal} {\bibinfo  {journal} {Light: Science \&
  Applications}\ }\textbf {\bibinfo {volume} {8}},\ \bibinfo {pages} {67}
  (\bibinfo {year} {2019})}\BibitemShut {NoStop}%
\bibitem [{\citenamefont {Chen}\ \emph {et~al.}(2020)\citenamefont {Chen},
  \citenamefont {Zhu},\ and\ \citenamefont {Capasso}}]{chen2020flat}%
  \BibitemOpen
  \bibfield  {author} {\bibinfo {author} {\bibfnamefont {W.~T.}\ \bibnamefont
  {Chen}}, \bibinfo {author} {\bibfnamefont {A.~Y.}\ \bibnamefont {Zhu}},\ and\
  \bibinfo {author} {\bibfnamefont {F.}~\bibnamefont {Capasso}},\ }\bibfield
  {title} {\bibinfo {title} {Flat optics with dispersion-engineered
  metasurfaces},\ }\href@noop {} {\bibfield  {journal} {\bibinfo  {journal}
  {Nature Reviews Materials}\ }\textbf {\bibinfo {volume} {5}},\ \bibinfo
  {pages} {604} (\bibinfo {year} {2020})}\BibitemShut {NoStop}%
\bibitem [{\citenamefont {Balli}\ \emph {et~al.}(2020)\citenamefont {Balli},
  \citenamefont {Sultan}, \citenamefont {Lami},\ and\ \citenamefont
  {Hastings}}]{balli2020hybrid}%
  \BibitemOpen
  \bibfield  {author} {\bibinfo {author} {\bibfnamefont {F.}~\bibnamefont
  {Balli}}, \bibinfo {author} {\bibfnamefont {M.}~\bibnamefont {Sultan}},
  \bibinfo {author} {\bibfnamefont {S.~K.}\ \bibnamefont {Lami}},\ and\
  \bibinfo {author} {\bibfnamefont {J.~T.}\ \bibnamefont {Hastings}},\
  }\bibfield  {title} {\bibinfo {title} {A hybrid achromatic metalens},\
  }\href@noop {} {\bibfield  {journal} {\bibinfo  {journal} {Nature
  communications}\ }\textbf {\bibinfo {volume} {11}},\ \bibinfo {pages} {3892}
  (\bibinfo {year} {2020})}\BibitemShut {NoStop}%
\bibitem [{\citenamefont {Chung}\ and\ \citenamefont
  {Miller}(2020)}]{chung2020high}%
  \BibitemOpen
  \bibfield  {author} {\bibinfo {author} {\bibfnamefont {H.}~\bibnamefont
  {Chung}}\ and\ \bibinfo {author} {\bibfnamefont {O.~D.}\ \bibnamefont
  {Miller}},\ }\bibfield  {title} {\bibinfo {title} {High-na achromatic
  metalenses by inverse design},\ }\href@noop {} {\bibfield  {journal}
  {\bibinfo  {journal} {Optics Express}\ }\textbf {\bibinfo {volume} {28}},\
  \bibinfo {pages} {6945} (\bibinfo {year} {2020})}\BibitemShut {NoStop}%
\bibitem [{\citenamefont {Li}\ \emph {et~al.}(2021{\natexlab{a}})\citenamefont
  {Li}, \citenamefont {Lin}, \citenamefont {Huang}, \citenamefont {Park},
  \citenamefont {Chen}, \citenamefont {Shi}, \citenamefont {Qiu}, \citenamefont
  {Cheng},\ and\ \citenamefont {Capasso}}]{li2021meta}%
  \BibitemOpen
  \bibfield  {author} {\bibinfo {author} {\bibfnamefont {Z.}~\bibnamefont
  {Li}}, \bibinfo {author} {\bibfnamefont {P.}~\bibnamefont {Lin}}, \bibinfo
  {author} {\bibfnamefont {Y.-W.}\ \bibnamefont {Huang}}, \bibinfo {author}
  {\bibfnamefont {J.-S.}\ \bibnamefont {Park}}, \bibinfo {author}
  {\bibfnamefont {W.~T.}\ \bibnamefont {Chen}}, \bibinfo {author}
  {\bibfnamefont {Z.}~\bibnamefont {Shi}}, \bibinfo {author} {\bibfnamefont
  {C.-W.}\ \bibnamefont {Qiu}}, \bibinfo {author} {\bibfnamefont {J.-X.}\
  \bibnamefont {Cheng}},\ and\ \bibinfo {author} {\bibfnamefont
  {F.}~\bibnamefont {Capasso}},\ }\bibfield  {title} {\bibinfo {title}
  {Meta-optics achieves rgb-achromatic focusing for virtual reality},\
  }\href@noop {} {\bibfield  {journal} {\bibinfo  {journal} {Science Advances}\
  }\textbf {\bibinfo {volume} {7}},\ \bibinfo {pages} {eabe4458} (\bibinfo
  {year} {2021}{\natexlab{a}})}\BibitemShut {NoStop}%
\bibitem [{\citenamefont {Li}\ \emph {et~al.}(2022)\citenamefont {Li},
  \citenamefont {Pestourie}, \citenamefont {Park}, \citenamefont {Huang},
  \citenamefont {Johnson},\ and\ \citenamefont {Capasso}}]{li2022inverse}%
  \BibitemOpen
  \bibfield  {author} {\bibinfo {author} {\bibfnamefont {Z.}~\bibnamefont
  {Li}}, \bibinfo {author} {\bibfnamefont {R.}~\bibnamefont {Pestourie}},
  \bibinfo {author} {\bibfnamefont {J.-S.}\ \bibnamefont {Park}}, \bibinfo
  {author} {\bibfnamefont {Y.-W.}\ \bibnamefont {Huang}}, \bibinfo {author}
  {\bibfnamefont {S.~G.}\ \bibnamefont {Johnson}},\ and\ \bibinfo {author}
  {\bibfnamefont {F.}~\bibnamefont {Capasso}},\ }\bibfield  {title} {\bibinfo
  {title} {Inverse design enables large-scale high-performance meta-optics
  reshaping virtual reality},\ }\href@noop {} {\bibfield  {journal} {\bibinfo
  {journal} {Nature communications}\ }\textbf {\bibinfo {volume} {13}},\
  \bibinfo {pages} {2409} (\bibinfo {year} {2022})}\BibitemShut {NoStop}%
\bibitem [{\citenamefont {Liu}\ \emph {et~al.}(2022)\citenamefont {Liu},
  \citenamefont {Zhang}, \citenamefont {Le~Roux}, \citenamefont {Cassan},
  \citenamefont {Marris-Morini}, \citenamefont {Vivien}, \citenamefont
  {Alonso-Ramos},\ and\ \citenamefont {Melati}}]{liu2022broadband}%
  \BibitemOpen
  \bibfield  {author} {\bibinfo {author} {\bibfnamefont {Y.}~\bibnamefont
  {Liu}}, \bibinfo {author} {\bibfnamefont {J.}~\bibnamefont {Zhang}}, \bibinfo
  {author} {\bibfnamefont {X.}~\bibnamefont {Le~Roux}}, \bibinfo {author}
  {\bibfnamefont {E.}~\bibnamefont {Cassan}}, \bibinfo {author} {\bibfnamefont
  {D.}~\bibnamefont {Marris-Morini}}, \bibinfo {author} {\bibfnamefont
  {L.}~\bibnamefont {Vivien}}, \bibinfo {author} {\bibfnamefont
  {C.}~\bibnamefont {Alonso-Ramos}},\ and\ \bibinfo {author} {\bibfnamefont
  {D.}~\bibnamefont {Melati}},\ }\bibfield  {title} {\bibinfo {title}
  {Broadband behavior of quadratic metalenses with a wide field of view},\
  }\href@noop {} {\bibfield  {journal} {\bibinfo  {journal} {Optics Express}\
  }\textbf {\bibinfo {volume} {30}},\ \bibinfo {pages} {39860} (\bibinfo {year}
  {2022})}\BibitemShut {NoStop}%
\bibitem [{\citenamefont {Sun}\ \emph {et~al.}(2022)\citenamefont {Sun},
  \citenamefont {Zhang}, \citenamefont {Dong}, \citenamefont {Feng},\ and\
  \citenamefont {Chu}}]{sun2022broadband}%
  \BibitemOpen
  \bibfield  {author} {\bibinfo {author} {\bibfnamefont {P.}~\bibnamefont
  {Sun}}, \bibinfo {author} {\bibfnamefont {M.}~\bibnamefont {Zhang}}, \bibinfo
  {author} {\bibfnamefont {F.}~\bibnamefont {Dong}}, \bibinfo {author}
  {\bibfnamefont {L.}~\bibnamefont {Feng}},\ and\ \bibinfo {author}
  {\bibfnamefont {W.}~\bibnamefont {Chu}},\ }\bibfield  {title} {\bibinfo
  {title} {Broadband achromatic polarization insensitive metalens over 950 nm
  bandwidth in the visible and near-infrared},\ }\href@noop {} {\bibfield
  {journal} {\bibinfo  {journal} {Chinese Optics Letters}\ }\textbf {\bibinfo
  {volume} {20}},\ \bibinfo {pages} {013601} (\bibinfo {year}
  {2022})}\BibitemShut {NoStop}%
\bibitem [{\citenamefont {Fan}\ \emph {et~al.}(2023)\citenamefont {Fan},
  \citenamefont {Yao},\ and\ \citenamefont {Tsai}}]{fan2023advance}%
  \BibitemOpen
  \bibfield  {author} {\bibinfo {author} {\bibfnamefont {Y.}~\bibnamefont
  {Fan}}, \bibinfo {author} {\bibfnamefont {J.}~\bibnamefont {Yao}},\ and\
  \bibinfo {author} {\bibfnamefont {D.~P.}\ \bibnamefont {Tsai}},\ }\bibfield
  {title} {\bibinfo {title} {Advance of large-area achromatic flat lenses},\
  }\href@noop {} {\bibfield  {journal} {\bibinfo  {journal} {Light: Science \&
  Applications}\ }\textbf {\bibinfo {volume} {12}},\ \bibinfo {pages} {51}
  (\bibinfo {year} {2023})}\BibitemShut {NoStop}%
\bibitem [{\citenamefont {Chu}\ \emph {et~al.}(2023)\citenamefont {Chu},
  \citenamefont {Xiao}, \citenamefont {Ye}, \citenamefont {Chen}, \citenamefont
  {Zhu},\ and\ \citenamefont {Li}}]{chu2023design}%
  \BibitemOpen
  \bibfield  {author} {\bibinfo {author} {\bibfnamefont {Y.}~\bibnamefont
  {Chu}}, \bibinfo {author} {\bibfnamefont {X.}~\bibnamefont {Xiao}}, \bibinfo
  {author} {\bibfnamefont {X.}~\bibnamefont {Ye}}, \bibinfo {author}
  {\bibfnamefont {C.}~\bibnamefont {Chen}}, \bibinfo {author} {\bibfnamefont
  {S.}~\bibnamefont {Zhu}},\ and\ \bibinfo {author} {\bibfnamefont
  {T.}~\bibnamefont {Li}},\ }\bibfield  {title} {\bibinfo {title} {Design of
  achromatic hybrid metalens with secondary spectrum correction},\ }\href@noop
  {} {\bibfield  {journal} {\bibinfo  {journal} {Optics Express}\ }\textbf
  {\bibinfo {volume} {31}},\ \bibinfo {pages} {21399} (\bibinfo {year}
  {2023})}\BibitemShut {NoStop}%
\bibitem [{\citenamefont {Hu}\ \emph {et~al.}(2023)\citenamefont {Hu},
  \citenamefont {Jiang}, \citenamefont {Zhang}, \citenamefont {Yang},
  \citenamefont {Ou}, \citenamefont {Li}, \citenamefont {Kong}, \citenamefont
  {Liu}, \citenamefont {Qiu},\ and\ \citenamefont {Duan}}]{hu2023asymptotic}%
  \BibitemOpen
  \bibfield  {author} {\bibinfo {author} {\bibfnamefont {Y.}~\bibnamefont
  {Hu}}, \bibinfo {author} {\bibfnamefont {Y.}~\bibnamefont {Jiang}}, \bibinfo
  {author} {\bibfnamefont {Y.}~\bibnamefont {Zhang}}, \bibinfo {author}
  {\bibfnamefont {X.}~\bibnamefont {Yang}}, \bibinfo {author} {\bibfnamefont
  {X.}~\bibnamefont {Ou}}, \bibinfo {author} {\bibfnamefont {L.}~\bibnamefont
  {Li}}, \bibinfo {author} {\bibfnamefont {X.}~\bibnamefont {Kong}}, \bibinfo
  {author} {\bibfnamefont {X.}~\bibnamefont {Liu}}, \bibinfo {author}
  {\bibfnamefont {C.-W.}\ \bibnamefont {Qiu}},\ and\ \bibinfo {author}
  {\bibfnamefont {H.}~\bibnamefont {Duan}},\ }\bibfield  {title} {\bibinfo
  {title} {Asymptotic dispersion engineering for ultra-broadband meta-optics},\
  }\href@noop {} {\bibfield  {journal} {\bibinfo  {journal} {nature
  communications}\ }\textbf {\bibinfo {volume} {14}},\ \bibinfo {pages} {6649}
  (\bibinfo {year} {2023})}\BibitemShut {NoStop}%
\bibitem [{\citenamefont {Pan}\ \emph {et~al.}(2023)\citenamefont {Pan},
  \citenamefont {Wang}, \citenamefont {Wang}, \citenamefont {S}, \citenamefont
  {Ruan}, \citenamefont {Wredh}, \citenamefont {Ke}, \citenamefont {Chan},
  \citenamefont {Zhang}, \citenamefont {Qiu} \emph {et~al.}}]{pan20233d}%
  \BibitemOpen
  \bibfield  {author} {\bibinfo {author} {\bibfnamefont {C.-F.}\ \bibnamefont
  {Pan}}, \bibinfo {author} {\bibfnamefont {H.}~\bibnamefont {Wang}}, \bibinfo
  {author} {\bibfnamefont {H.}~\bibnamefont {Wang}}, \bibinfo {author}
  {\bibfnamefont {P.~N.}\ \bibnamefont {S}}, \bibinfo {author} {\bibfnamefont
  {Q.}~\bibnamefont {Ruan}}, \bibinfo {author} {\bibfnamefont {S.}~\bibnamefont
  {Wredh}}, \bibinfo {author} {\bibfnamefont {Y.}~\bibnamefont {Ke}}, \bibinfo
  {author} {\bibfnamefont {J.~Y.~E.}\ \bibnamefont {Chan}}, \bibinfo {author}
  {\bibfnamefont {W.}~\bibnamefont {Zhang}}, \bibinfo {author} {\bibfnamefont
  {C.-W.}\ \bibnamefont {Qiu}}, \emph {et~al.},\ }\bibfield  {title} {\bibinfo
  {title} {3d-printed multilayer structures for high--numerical aperture
  achromatic metalenses},\ }\href@noop {} {\bibfield  {journal} {\bibinfo
  {journal} {Science advances}\ }\textbf {\bibinfo {volume} {9}},\ \bibinfo
  {pages} {eadj9262} (\bibinfo {year} {2023})}\BibitemShut {NoStop}%
\bibitem [{\citenamefont {Pan}\ \emph {et~al.}(2022)\citenamefont {Pan},
  \citenamefont {Fu}, \citenamefont {Zheng}, \citenamefont {Chen},
  \citenamefont {Zang}, \citenamefont {Duan}, \citenamefont {Li}, \citenamefont
  {Qiu},\ and\ \citenamefont {Hu}}]{pan2022dielectric}%
  \BibitemOpen
  \bibfield  {author} {\bibinfo {author} {\bibfnamefont {M.}~\bibnamefont
  {Pan}}, \bibinfo {author} {\bibfnamefont {Y.}~\bibnamefont {Fu}}, \bibinfo
  {author} {\bibfnamefont {M.}~\bibnamefont {Zheng}}, \bibinfo {author}
  {\bibfnamefont {H.}~\bibnamefont {Chen}}, \bibinfo {author} {\bibfnamefont
  {Y.}~\bibnamefont {Zang}}, \bibinfo {author} {\bibfnamefont {H.}~\bibnamefont
  {Duan}}, \bibinfo {author} {\bibfnamefont {Q.}~\bibnamefont {Li}}, \bibinfo
  {author} {\bibfnamefont {M.}~\bibnamefont {Qiu}},\ and\ \bibinfo {author}
  {\bibfnamefont {Y.}~\bibnamefont {Hu}},\ }\bibfield  {title} {\bibinfo
  {title} {Dielectric metalens for miniaturized imaging systems: progress and
  challenges},\ }\href@noop {} {\bibfield  {journal} {\bibinfo  {journal}
  {Light: Science \& Applications}\ }\textbf {\bibinfo {volume} {11}},\
  \bibinfo {pages} {195} (\bibinfo {year} {2022})}\BibitemShut {NoStop}%
\bibitem [{\citenamefont {Liang}\ \emph {et~al.}(2019)\citenamefont {Liang},
  \citenamefont {Martins}, \citenamefont {Borges}, \citenamefont {Zhou},
  \citenamefont {Martins}, \citenamefont {Li},\ and\ \citenamefont
  {Krauss}}]{liang2019high}%
  \BibitemOpen
  \bibfield  {author} {\bibinfo {author} {\bibfnamefont {H.}~\bibnamefont
  {Liang}}, \bibinfo {author} {\bibfnamefont {A.}~\bibnamefont {Martins}},
  \bibinfo {author} {\bibfnamefont {B.-H.~V.}\ \bibnamefont {Borges}}, \bibinfo
  {author} {\bibfnamefont {J.}~\bibnamefont {Zhou}}, \bibinfo {author}
  {\bibfnamefont {E.~R.}\ \bibnamefont {Martins}}, \bibinfo {author}
  {\bibfnamefont {J.}~\bibnamefont {Li}},\ and\ \bibinfo {author}
  {\bibfnamefont {T.~F.}\ \bibnamefont {Krauss}},\ }\bibfield  {title}
  {\bibinfo {title} {High performance metalenses: numerical aperture,
  aberrations, chromaticity, and trade-offs},\ }\href@noop {} {\bibfield
  {journal} {\bibinfo  {journal} {Optica}\ }\textbf {\bibinfo {volume} {6}},\
  \bibinfo {pages} {1461} (\bibinfo {year} {2019})}\BibitemShut {NoStop}%
\bibitem [{\citenamefont {Lassalle}\ \emph {et~al.}(2021)\citenamefont
  {Lassalle}, \citenamefont {Mass}, \citenamefont {Eschimese}, \citenamefont
  {Baranikov}, \citenamefont {Khaidarov}, \citenamefont {Li}, \citenamefont
  {Paniagua-Dominguez},\ and\ \citenamefont {Kuznetsov}}]{lassalle2021imaging}%
  \BibitemOpen
  \bibfield  {author} {\bibinfo {author} {\bibfnamefont {E.}~\bibnamefont
  {Lassalle}}, \bibinfo {author} {\bibfnamefont {T.~W.~W.}\ \bibnamefont
  {Mass}}, \bibinfo {author} {\bibfnamefont {D.}~\bibnamefont {Eschimese}},
  \bibinfo {author} {\bibfnamefont {A.~V.}\ \bibnamefont {Baranikov}}, \bibinfo
  {author} {\bibfnamefont {E.}~\bibnamefont {Khaidarov}}, \bibinfo {author}
  {\bibfnamefont {S.}~\bibnamefont {Li}}, \bibinfo {author} {\bibfnamefont
  {R.}~\bibnamefont {Paniagua-Dominguez}},\ and\ \bibinfo {author}
  {\bibfnamefont {A.~I.}\ \bibnamefont {Kuznetsov}},\ }\bibfield  {title}
  {\bibinfo {title} {Imaging properties of large field-of-view quadratic
  metalenses and their applications to fingerprint detection},\ }\href
  {https://doi.org/10.1021/acsphotonics.1c00237} {\bibfield  {journal}
  {\bibinfo  {journal} {ACS Photonics}\ }\textbf {\bibinfo {volume} {8}},\
  \bibinfo {pages} {1457} (\bibinfo {year} {2021})}\BibitemShut {NoStop}%
\bibitem [{\citenamefont {Arbabi}\ \emph {et~al.}(2017)\citenamefont {Arbabi},
  \citenamefont {Arbabi}, \citenamefont {Horie}, \citenamefont {Kamali},\ and\
  \citenamefont {Faraon}}]{arbabi2017planar}%
  \BibitemOpen
  \bibfield  {author} {\bibinfo {author} {\bibfnamefont {A.}~\bibnamefont
  {Arbabi}}, \bibinfo {author} {\bibfnamefont {E.}~\bibnamefont {Arbabi}},
  \bibinfo {author} {\bibfnamefont {Y.}~\bibnamefont {Horie}}, \bibinfo
  {author} {\bibfnamefont {S.~M.}\ \bibnamefont {Kamali}},\ and\ \bibinfo
  {author} {\bibfnamefont {A.}~\bibnamefont {Faraon}},\ }\bibfield  {title}
  {\bibinfo {title} {Planar metasurface retroreflector},\ }\href@noop {}
  {\bibfield  {journal} {\bibinfo  {journal} {Nature Photonics}\ }\textbf
  {\bibinfo {volume} {11}},\ \bibinfo {pages} {415} (\bibinfo {year}
  {2017})}\BibitemShut {NoStop}%
\bibitem [{\citenamefont {Groever}\ \emph {et~al.}(2017)\citenamefont
  {Groever}, \citenamefont {Chen},\ and\ \citenamefont
  {Capasso}}]{groever2017meta}%
  \BibitemOpen
  \bibfield  {author} {\bibinfo {author} {\bibfnamefont {B.}~\bibnamefont
  {Groever}}, \bibinfo {author} {\bibfnamefont {W.~T.}\ \bibnamefont {Chen}},\
  and\ \bibinfo {author} {\bibfnamefont {F.}~\bibnamefont {Capasso}},\
  }\bibfield  {title} {\bibinfo {title} {Meta-lens doublet in the visible
  region},\ }\href@noop {} {\bibfield  {journal} {\bibinfo  {journal} {Nano
  letters}\ }\textbf {\bibinfo {volume} {17}},\ \bibinfo {pages} {4902}
  (\bibinfo {year} {2017})}\BibitemShut {NoStop}%
\bibitem [{\citenamefont {Pu}\ \emph {et~al.}(2017)\citenamefont {Pu},
  \citenamefont {Li}, \citenamefont {Guo}, \citenamefont {Ma},\ and\
  \citenamefont {Luo}}]{pu2017nanoapertures}%
  \BibitemOpen
  \bibfield  {author} {\bibinfo {author} {\bibfnamefont {M.}~\bibnamefont
  {Pu}}, \bibinfo {author} {\bibfnamefont {X.}~\bibnamefont {Li}}, \bibinfo
  {author} {\bibfnamefont {Y.}~\bibnamefont {Guo}}, \bibinfo {author}
  {\bibfnamefont {X.}~\bibnamefont {Ma}},\ and\ \bibinfo {author}
  {\bibfnamefont {X.}~\bibnamefont {Luo}},\ }\bibfield  {title} {\bibinfo
  {title} {Nanoapertures with ordered rotations: symmetry transformation and
  wide-angle flat lensing},\ }\href@noop {} {\bibfield  {journal} {\bibinfo
  {journal} {Optics Express}\ }\textbf {\bibinfo {volume} {25}},\ \bibinfo
  {pages} {31471} (\bibinfo {year} {2017})}\BibitemShut {NoStop}%
\bibitem [{\citenamefont {Martins}\ \emph {et~al.}(2020)\citenamefont
  {Martins}, \citenamefont {Li}, \citenamefont {Li}, \citenamefont {Liang},
  \citenamefont {Conteduca}, \citenamefont {Borges}, \citenamefont {Krauss},\
  and\ \citenamefont {Martins}}]{martins2020metalenses}%
  \BibitemOpen
  \bibfield  {author} {\bibinfo {author} {\bibfnamefont {A.}~\bibnamefont
  {Martins}}, \bibinfo {author} {\bibfnamefont {K.}~\bibnamefont {Li}},
  \bibinfo {author} {\bibfnamefont {J.}~\bibnamefont {Li}}, \bibinfo {author}
  {\bibfnamefont {H.}~\bibnamefont {Liang}}, \bibinfo {author} {\bibfnamefont
  {D.}~\bibnamefont {Conteduca}}, \bibinfo {author} {\bibfnamefont {B.-H.~V.}\
  \bibnamefont {Borges}}, \bibinfo {author} {\bibfnamefont {T.~F.}\
  \bibnamefont {Krauss}},\ and\ \bibinfo {author} {\bibfnamefont {E.~R.}\
  \bibnamefont {Martins}},\ }\bibfield  {title} {\bibinfo {title} {On
  metalenses with arbitrarily wide field of view},\ }\href@noop {} {\bibfield
  {journal} {\bibinfo  {journal} {Acs Photonics}\ }\textbf {\bibinfo {volume}
  {7}},\ \bibinfo {pages} {2073} (\bibinfo {year} {2020})}\BibitemShut
  {NoStop}%
\bibitem [{\citenamefont {Engelberg}\ \emph {et~al.}(2020)\citenamefont
  {Engelberg}, \citenamefont {Zhou}, \citenamefont {Mazurski}, \citenamefont
  {Bar-David}, \citenamefont {Kristensen},\ and\ \citenamefont
  {Levy}}]{engelberg2020near}%
  \BibitemOpen
  \bibfield  {author} {\bibinfo {author} {\bibfnamefont {J.}~\bibnamefont
  {Engelberg}}, \bibinfo {author} {\bibfnamefont {C.}~\bibnamefont {Zhou}},
  \bibinfo {author} {\bibfnamefont {N.}~\bibnamefont {Mazurski}}, \bibinfo
  {author} {\bibfnamefont {J.}~\bibnamefont {Bar-David}}, \bibinfo {author}
  {\bibfnamefont {A.}~\bibnamefont {Kristensen}},\ and\ \bibinfo {author}
  {\bibfnamefont {U.}~\bibnamefont {Levy}},\ }\bibfield  {title} {\bibinfo
  {title} {Near-ir wide-field-of-view huygens metalens for outdoor imaging
  applications},\ }\href@noop {} {\bibfield  {journal} {\bibinfo  {journal}
  {Nanophotonics}\ }\textbf {\bibinfo {volume} {9}},\ \bibinfo {pages} {361}
  (\bibinfo {year} {2020})}\BibitemShut {NoStop}%
\bibitem [{\citenamefont {Chen}\ \emph {et~al.}(2022)\citenamefont {Chen},
  \citenamefont {Ye}, \citenamefont {Gao}, \citenamefont {Chen}, \citenamefont
  {Zhao}, \citenamefont {Huang}, \citenamefont {Qiu}, \citenamefont {Zhu},\
  and\ \citenamefont {Li}}]{chen2022planar}%
  \BibitemOpen
  \bibfield  {author} {\bibinfo {author} {\bibfnamefont {J.}~\bibnamefont
  {Chen}}, \bibinfo {author} {\bibfnamefont {X.}~\bibnamefont {Ye}}, \bibinfo
  {author} {\bibfnamefont {S.}~\bibnamefont {Gao}}, \bibinfo {author}
  {\bibfnamefont {Y.}~\bibnamefont {Chen}}, \bibinfo {author} {\bibfnamefont
  {Y.}~\bibnamefont {Zhao}}, \bibinfo {author} {\bibfnamefont {C.}~\bibnamefont
  {Huang}}, \bibinfo {author} {\bibfnamefont {K.}~\bibnamefont {Qiu}}, \bibinfo
  {author} {\bibfnamefont {S.}~\bibnamefont {Zhu}},\ and\ \bibinfo {author}
  {\bibfnamefont {T.}~\bibnamefont {Li}},\ }\bibfield  {title} {\bibinfo
  {title} {Planar wide-angle-imaging camera enabled by metalens array},\
  }\href@noop {} {\bibfield  {journal} {\bibinfo  {journal} {Optica}\ }\textbf
  {\bibinfo {volume} {9}},\ \bibinfo {pages} {431} (\bibinfo {year}
  {2022})}\BibitemShut {NoStop}%
\bibitem [{\citenamefont {Fan}\ \emph {et~al.}(2020)\citenamefont {Fan},
  \citenamefont {Lin},\ and\ \citenamefont {Su}}]{fan2020ultrawide}%
  \BibitemOpen
  \bibfield  {author} {\bibinfo {author} {\bibfnamefont {C.-Y.}\ \bibnamefont
  {Fan}}, \bibinfo {author} {\bibfnamefont {C.-P.}\ \bibnamefont {Lin}},\ and\
  \bibinfo {author} {\bibfnamefont {G.-D.~J.}\ \bibnamefont {Su}},\ }\bibfield
  {title} {\bibinfo {title} {Ultrawide-angle and high-efficiency metalens in
  hexagonal arrangement},\ }\href@noop {} {\bibfield  {journal} {\bibinfo
  {journal} {Scientific Reports}\ }\textbf {\bibinfo {volume} {10}},\ \bibinfo
  {pages} {15677} (\bibinfo {year} {2020})}\BibitemShut {NoStop}%
\bibitem [{\citenamefont {Li}\ \emph {et~al.}(2021{\natexlab{b}})\citenamefont
  {Li}, \citenamefont {Wang}, \citenamefont {Wang}, \citenamefont {Lu},
  \citenamefont {Guo}, \citenamefont {Li}, \citenamefont {Ma}, \citenamefont
  {Pu},\ and\ \citenamefont {Luo}}]{li2021super}%
  \BibitemOpen
  \bibfield  {author} {\bibinfo {author} {\bibfnamefont {Z.}~\bibnamefont
  {Li}}, \bibinfo {author} {\bibfnamefont {C.}~\bibnamefont {Wang}}, \bibinfo
  {author} {\bibfnamefont {Y.}~\bibnamefont {Wang}}, \bibinfo {author}
  {\bibfnamefont {X.}~\bibnamefont {Lu}}, \bibinfo {author} {\bibfnamefont
  {Y.}~\bibnamefont {Guo}}, \bibinfo {author} {\bibfnamefont {X.}~\bibnamefont
  {Li}}, \bibinfo {author} {\bibfnamefont {X.}~\bibnamefont {Ma}}, \bibinfo
  {author} {\bibfnamefont {M.}~\bibnamefont {Pu}},\ and\ \bibinfo {author}
  {\bibfnamefont {X.}~\bibnamefont {Luo}},\ }\bibfield  {title} {\bibinfo
  {title} {Super-oscillatory metasurface doublet for sub-diffraction focusing
  with a large incident angle},\ }\href@noop {} {\bibfield  {journal} {\bibinfo
   {journal} {Optics Express}\ }\textbf {\bibinfo {volume} {29}},\ \bibinfo
  {pages} {9991} (\bibinfo {year} {2021}{\natexlab{b}})}\BibitemShut {NoStop}%
\bibitem [{\citenamefont {Yang}\ \emph {et~al.}(2023)\citenamefont {Yang},
  \citenamefont {Shalaginov}, \citenamefont {Lin}, \citenamefont {An},
  \citenamefont {Agarwal}, \citenamefont {Zhang}, \citenamefont
  {Rivero-Baleine}, \citenamefont {Gu},\ and\ \citenamefont
  {Hu}}]{yang2023wide}%
  \BibitemOpen
  \bibfield  {author} {\bibinfo {author} {\bibfnamefont {F.}~\bibnamefont
  {Yang}}, \bibinfo {author} {\bibfnamefont {M.~Y.}\ \bibnamefont
  {Shalaginov}}, \bibinfo {author} {\bibfnamefont {H.-I.}\ \bibnamefont {Lin}},
  \bibinfo {author} {\bibfnamefont {S.}~\bibnamefont {An}}, \bibinfo {author}
  {\bibfnamefont {A.}~\bibnamefont {Agarwal}}, \bibinfo {author} {\bibfnamefont
  {H.}~\bibnamefont {Zhang}}, \bibinfo {author} {\bibfnamefont
  {C.}~\bibnamefont {Rivero-Baleine}}, \bibinfo {author} {\bibfnamefont
  {T.}~\bibnamefont {Gu}},\ and\ \bibinfo {author} {\bibfnamefont
  {J.}~\bibnamefont {Hu}},\ }\bibfield  {title} {\bibinfo {title} {Wide
  field-of-view metalens: a tutorial},\ }\href@noop {} {\bibfield  {journal}
  {\bibinfo  {journal} {Advanced Photonics}\ }\textbf {\bibinfo {volume} {5}},\
  \bibinfo {pages} {033001} (\bibinfo {year} {2023})}\BibitemShut {NoStop}%
\bibitem [{\citenamefont {Xie}\ \emph {et~al.}(2023)\citenamefont {Xie},
  \citenamefont {Carson}, \citenamefont {Fr{\"o}ch}, \citenamefont {Majumdar},
  \citenamefont {Seibel},\ and\ \citenamefont {B{\"o}hringer}}]{xie2023large}%
  \BibitemOpen
  \bibfield  {author} {\bibinfo {author} {\bibfnamefont {N.}~\bibnamefont
  {Xie}}, \bibinfo {author} {\bibfnamefont {M.~D.}\ \bibnamefont {Carson}},
  \bibinfo {author} {\bibfnamefont {J.~E.}\ \bibnamefont {Fr{\"o}ch}}, \bibinfo
  {author} {\bibfnamefont {A.}~\bibnamefont {Majumdar}}, \bibinfo {author}
  {\bibfnamefont {E.~J.}\ \bibnamefont {Seibel}},\ and\ \bibinfo {author}
  {\bibfnamefont {K.~F.}\ \bibnamefont {B{\"o}hringer}},\ }\bibfield  {title}
  {\bibinfo {title} {Large field-of-view short-wave infrared metalens for
  scanning fiber endoscopy},\ }\href@noop {} {\bibfield  {journal} {\bibinfo
  {journal} {Journal of Biomedical Optics}\ }\textbf {\bibinfo {volume} {28}},\
  \bibinfo {pages} {094802} (\bibinfo {year} {2023})}\BibitemShut {NoStop}%
\bibitem [{\citenamefont {Dong}\ \emph {et~al.}(2025)\citenamefont {Dong},
  \citenamefont {Zheng}, \citenamefont {Yang}, \citenamefont {Tang},
  \citenamefont {Zhao}, \citenamefont {Huang}, \citenamefont {Gu},
  \citenamefont {Hu},\ and\ \citenamefont {Zhang}}]{dong2025full}%
  \BibitemOpen
  \bibfield  {author} {\bibinfo {author} {\bibfnamefont {Y.}~\bibnamefont
  {Dong}}, \bibinfo {author} {\bibfnamefont {B.}~\bibnamefont {Zheng}},
  \bibinfo {author} {\bibfnamefont {F.}~\bibnamefont {Yang}}, \bibinfo {author}
  {\bibfnamefont {H.}~\bibnamefont {Tang}}, \bibinfo {author} {\bibfnamefont
  {H.}~\bibnamefont {Zhao}}, \bibinfo {author} {\bibfnamefont {Y.}~\bibnamefont
  {Huang}}, \bibinfo {author} {\bibfnamefont {T.}~\bibnamefont {Gu}}, \bibinfo
  {author} {\bibfnamefont {J.}~\bibnamefont {Hu}},\ and\ \bibinfo {author}
  {\bibfnamefont {H.}~\bibnamefont {Zhang}},\ }\bibfield  {title} {\bibinfo
  {title} {Full-color, wide field-of-view metalens imaging via deep learning},\
  }\href@noop {} {\bibfield  {journal} {\bibinfo  {journal} {Advanced Optical
  Materials}\ }\textbf {\bibinfo {volume} {13}},\ \bibinfo {pages} {2402207}
  (\bibinfo {year} {2025})}\BibitemShut {NoStop}%
\bibitem [{\citenamefont {Shalaginov}\ \emph {et~al.}(2020)\citenamefont
  {Shalaginov}, \citenamefont {An}, \citenamefont {Yang}, \citenamefont {Su},
  \citenamefont {Lyzwa}, \citenamefont {Agarwal}, \citenamefont {Zhang},
  \citenamefont {Hu},\ and\ \citenamefont {Gu}}]{shalaginov2020single}%
  \BibitemOpen
  \bibfield  {author} {\bibinfo {author} {\bibfnamefont {M.~Y.}\ \bibnamefont
  {Shalaginov}}, \bibinfo {author} {\bibfnamefont {S.}~\bibnamefont {An}},
  \bibinfo {author} {\bibfnamefont {F.}~\bibnamefont {Yang}}, \bibinfo {author}
  {\bibfnamefont {P.}~\bibnamefont {Su}}, \bibinfo {author} {\bibfnamefont
  {D.}~\bibnamefont {Lyzwa}}, \bibinfo {author} {\bibfnamefont {A.~M.}\
  \bibnamefont {Agarwal}}, \bibinfo {author} {\bibfnamefont {H.}~\bibnamefont
  {Zhang}}, \bibinfo {author} {\bibfnamefont {J.}~\bibnamefont {Hu}},\ and\
  \bibinfo {author} {\bibfnamefont {T.}~\bibnamefont {Gu}},\ }\bibfield
  {title} {\bibinfo {title} {Single-element diffraction-limited fisheye
  metalens},\ }\href@noop {} {\bibfield  {journal} {\bibinfo  {journal} {Nano
  Letters}\ }\textbf {\bibinfo {volume} {20}},\ \bibinfo {pages} {7429}
  (\bibinfo {year} {2020})}\BibitemShut {NoStop}%
\bibitem [{\citenamefont {Arbabi}\ \emph {et~al.}(2016)\citenamefont {Arbabi},
  \citenamefont {Arbabi}, \citenamefont {Kamali}, \citenamefont {Horie},
  \citenamefont {Han},\ and\ \citenamefont {Faraon}}]{arbabi2016miniature}%
  \BibitemOpen
  \bibfield  {author} {\bibinfo {author} {\bibfnamefont {A.}~\bibnamefont
  {Arbabi}}, \bibinfo {author} {\bibfnamefont {E.}~\bibnamefont {Arbabi}},
  \bibinfo {author} {\bibfnamefont {S.~M.}\ \bibnamefont {Kamali}}, \bibinfo
  {author} {\bibfnamefont {Y.}~\bibnamefont {Horie}}, \bibinfo {author}
  {\bibfnamefont {S.}~\bibnamefont {Han}},\ and\ \bibinfo {author}
  {\bibfnamefont {A.}~\bibnamefont {Faraon}},\ }\bibfield  {title} {\bibinfo
  {title} {Miniature optical planar camera based on a wide-angle metasurface
  doublet corrected for monochromatic aberrations},\ }\href@noop {} {\bibfield
  {journal} {\bibinfo  {journal} {Nature communications}\ }\textbf {\bibinfo
  {volume} {7}},\ \bibinfo {pages} {13682} (\bibinfo {year}
  {2016})}\BibitemShut {NoStop}%
\bibitem [{\citenamefont {Xu}\ \emph {et~al.}(2023)\citenamefont {Xu},
  \citenamefont {Chen}, \citenamefont {Li}, \citenamefont {Liu},\ and\
  \citenamefont {Chen}}]{xu2023broadband}%
  \BibitemOpen
  \bibfield  {author} {\bibinfo {author} {\bibfnamefont {F.}~\bibnamefont
  {Xu}}, \bibinfo {author} {\bibfnamefont {W.}~\bibnamefont {Chen}}, \bibinfo
  {author} {\bibfnamefont {M.}~\bibnamefont {Li}}, \bibinfo {author}
  {\bibfnamefont {P.}~\bibnamefont {Liu}},\ and\ \bibinfo {author}
  {\bibfnamefont {Y.}~\bibnamefont {Chen}},\ }\bibfield  {title} {\bibinfo
  {title} {Broadband achromatic and wide field-of-view single-layer metalenses
  in the mid-infrared},\ }\href@noop {} {\bibfield  {journal} {\bibinfo
  {journal} {Optics Express}\ }\textbf {\bibinfo {volume} {31}},\ \bibinfo
  {pages} {36439} (\bibinfo {year} {2023})}\BibitemShut {NoStop}%
\bibitem [{\citenamefont {Hongli}\ \emph {et~al.}(2024)\citenamefont {Hongli},
  \citenamefont {Zhaofeng},\ and\ \citenamefont
  {Xiaotong}}]{hongli2024broadband}%
  \BibitemOpen
  \bibfield  {author} {\bibinfo {author} {\bibfnamefont {Y.}~\bibnamefont
  {Hongli}}, \bibinfo {author} {\bibfnamefont {C.}~\bibnamefont {Zhaofeng}},\
  and\ \bibinfo {author} {\bibfnamefont {L.}~\bibnamefont {Xiaotong}},\
  }\bibfield  {title} {\bibinfo {title} {Broadband achromatic and wide field of
  view metalens-doublet by inverse design},\ }\href@noop {} {\bibfield
  {journal} {\bibinfo  {journal} {Optics Express}\ }\textbf {\bibinfo {volume}
  {32}},\ \bibinfo {pages} {15315} (\bibinfo {year} {2024})}\BibitemShut
  {NoStop}%
\bibitem [{\citenamefont {Yang}\ \emph {et~al.}(2021)\citenamefont {Yang},
  \citenamefont {An}, \citenamefont {Shalaginov}, \citenamefont {Zhang},
  \citenamefont {Rivero-Baleine}, \citenamefont {Hu},\ and\ \citenamefont
  {Gu}}]{Yang:21}%
  \BibitemOpen
  \bibfield  {author} {\bibinfo {author} {\bibfnamefont {F.}~\bibnamefont
  {Yang}}, \bibinfo {author} {\bibfnamefont {S.}~\bibnamefont {An}}, \bibinfo
  {author} {\bibfnamefont {M.~Y.}\ \bibnamefont {Shalaginov}}, \bibinfo
  {author} {\bibfnamefont {H.}~\bibnamefont {Zhang}}, \bibinfo {author}
  {\bibfnamefont {C.}~\bibnamefont {Rivero-Baleine}}, \bibinfo {author}
  {\bibfnamefont {J.}~\bibnamefont {Hu}},\ and\ \bibinfo {author}
  {\bibfnamefont {T.}~\bibnamefont {Gu}},\ }\bibfield  {title} {\bibinfo
  {title} {Design of broadband and wide-field-of-view metalenses},\ }\href
  {https://doi.org/10.1364/OL.439393} {\bibfield  {journal} {\bibinfo
  {journal} {Opt. Lett.}\ }\textbf {\bibinfo {volume} {46}},\ \bibinfo {pages}
  {5735} (\bibinfo {year} {2021})}\BibitemShut {NoStop}%
\bibitem [{\citenamefont {Hugonin}\ and\ \citenamefont
  {Lalanne}(2021)}]{hugonin2021reticolo}%
  \BibitemOpen
  \bibfield  {author} {\bibinfo {author} {\bibfnamefont {J.~P.}\ \bibnamefont
  {Hugonin}}\ and\ \bibinfo {author} {\bibfnamefont {P.}~\bibnamefont
  {Lalanne}},\ }\bibfield  {title} {\bibinfo {title} {Reticolo software for
  grating analysis},\ }\href@noop {} {\bibfield  {journal} {\bibinfo  {journal}
  {arXiv preprint arXiv:2101.00901}\ } (\bibinfo {year} {2021})}\BibitemShut
  {NoStop}%
\bibitem [{\citenamefont {Ueno}\ \emph {et~al.}(2024)\citenamefont {Ueno},
  \citenamefont {Hu},\ and\ \citenamefont {An}}]{ueno2024ai}%
  \BibitemOpen
  \bibfield  {author} {\bibinfo {author} {\bibfnamefont {A.}~\bibnamefont
  {Ueno}}, \bibinfo {author} {\bibfnamefont {J.}~\bibnamefont {Hu}},\ and\
  \bibinfo {author} {\bibfnamefont {S.}~\bibnamefont {An}},\ }\bibfield
  {title} {\bibinfo {title} {Ai for optical metasurface},\ }\href@noop {}
  {\bibfield  {journal} {\bibinfo  {journal} {npj Nanophotonics}\ }\textbf
  {\bibinfo {volume} {1}},\ \bibinfo {pages} {36} (\bibinfo {year}
  {2024})}\BibitemShut {NoStop}%
\bibitem [{\citenamefont {Presutti}\ and\ \citenamefont
  {Monticone}(2020)}]{presutti2020focusing}%
  \BibitemOpen
  \bibfield  {author} {\bibinfo {author} {\bibfnamefont {F.}~\bibnamefont
  {Presutti}}\ and\ \bibinfo {author} {\bibfnamefont {F.}~\bibnamefont
  {Monticone}},\ }\bibfield  {title} {\bibinfo {title} {Focusing on bandwidth:
  achromatic metalens limits},\ }\href@noop {} {\bibfield  {journal} {\bibinfo
  {journal} {Optica}\ }\textbf {\bibinfo {volume} {7}},\ \bibinfo {pages} {624}
  (\bibinfo {year} {2020})}\BibitemShut {NoStop}%
\end{thebibliography}%

\end{document}